\documentclass[letterpaper,journal,10pt]{IEEEtran}
\usepackage{graphicx}
\usepackage{epstopdf}
\usepackage{rotating}

\usepackage{cite}
\usepackage{url}

\usepackage[cmex10]{amsmath}
\usepackage{amssymb,amsfonts,bm}

\usepackage[font=small]{caption}
\usepackage[subrefformat=parens,labelformat=parens,skip=0pt]{subcaption}

\usepackage{calc}

\usepackage[usenames]{color}
\usepackage[normalem]{ulem}  

\DeclareMathOperator*{\argmin}{arg\,min}
\DeclareMathOperator*{\argmax}{arg\,max}
\DeclareMathOperator{\Poisson}{Poisson}
\DeclareMathOperator{\binomial}{binomial}
\DeclareMathOperator{\pen}{pen}

\def\alphahatPML{\widehat{\alpha}^{\rm PML}}
\def\balpha{\bm{\alpha}}
\def\bd{b_{\rm d}}
\def\bz{\bm{z}}
\def\dsp{d_{\rm sp}}
\def\dspmax{d_{\rm sp}^{\rm max}}
\def\ktilde{\tilde{k}}
\def\lambdab{\lambda^{\rm b}}
\def\lambdas{\lambda_{i,j}^{\rm s}}

\def\Ncl{N_{\rm cl}}
\def\Neighborhood{\mathcal{N}}
\def\Nmax{k_{i,j}^{\rm max}}
\def\Nr{N_{\rm r}}
\def\Nsp{N^{\rm sp}_{i,j}}

\def\Pwind{P_{\rm wind}}
\def\R{\mathbb{R}}
\def\ShinEtAl{Shin, \emph{et al.}}
\def\Shin{\ShinEtAl~\cite{Shin2015}}
\def\tauFA{\tau_{\rm FA}}
\def\tausp{\tau_{\rm sp}}
\def\Tp{T_{\rm p}}
\def\Tr{T_{\rm r}}
\def\ttilde{\tilde{t}}
\def\Twind{T_{\rm wind}}
\def\zmax{z_{\rm max}}
\def\MSE{{\rm MSE}}
\def\RMSE{{\rm RMSE}}

\title{A Few Photons Among Many: Unmixing Signal and 
Noise for Photon-Efficient Active Imaging
\thanks{This material is based upon work supported
in part by a Draper Fellowship
and by the US National Science Foundation under Grant No.\ 1422034.}
\thanks{The authors are with the Department of Electrical and Computer Engineering, Boston University, Boston, MA 02215 USA.}
}

\author{Joshua~Rapp,~\IEEEmembership{Student Member,~IEEE}
        and Vivek~K~Goyal,~\IEEEmembership{Fellow,~IEEE}}%

\begin{document}

\maketitle

\begin{abstract}
Conventional LIDAR systems require hundreds or thousands of photon detections to form accurate depth and reflectivity images. 
Recent photon-efficient computational imaging methods are remarkably effective with only 1.0 to 3.0 detected photons per pixel,
but they are not demonstrated at signal-to-background ratio (SBR) below 1.0 because their imaging accuracies degrade significantly in the presence of high background noise.
We introduce a new approach to depth and reflectivity estimation that focuses on unmixing contributions from signal and noise sources.
At each pixel in an image, short-duration range gates are adaptively determined and applied to remove detections likely to be due to noise. 
For pixels with too few detections to perform this censoring accurately, we borrow data from neighboring pixels to improve depth estimates,
where the neighborhood formation is also adaptive to scene content. 
Algorithm performance is demonstrated on experimental data at varying levels of noise. 
Results show improved performance of both reflectivity and depth estimates over state-of-the-art methods, especially at low signal-to-background ratios.
In particular, accurate imaging is demonstrated with SBR as low as 0.04\@.
This validation of a photon-efficient, noise-tolerant method demonstrates the viability of rapid, long-range, and low-power LIDAR imaging.
\end{abstract}

\section{Introduction}
Non-contact depth measurement has a wide range of uses, from industrial to military or scientific purposes.
Active optical methods, such as light detection and ranging 
(LIDAR) systems, are especially useful due to their high spatial resolution
relative to RADAR or ultrasound methods~\cite{Amann2001}. 
As a result, LIDAR has been successfully used for applications as varied as forest biomass estimation~\cite{Popescu2007},
geological surveying~\cite{Stoker2008}, land mine detection~\cite{Shaw2005}, and autonomous navigation~\cite{Zhang2010,Wooden2010}. 
LIDAR systems have recently begun to employ single-photon avalanche diode (SPAD) detectors as sensors, replacing the previously used photomultiplier tubes (PMTs).
These SPAD detectors allow for measurement of signals with much lower intensities,
such as those from distant, poorly reflective, or oblique-angled surfaces~\cite{Pellegrini2000}, or power-limited systems used for covert imaging or in mobile applications~\cite{ColacoKYGSG:13}.
By forming histograms from hundreds to thousands of repeated measurements,
photon-counting systems can approximate the full-waveform output of a PMT
and use cross-correlation or maximum likelihood (ML) estimation to determine the depth of a scene~\cite{McCarthy2009,Wallace2006}.

New photon-counting LIDAR systems have demonstrated dramatic improvements in photon efficiency, forming accurate depth and reflectivity images from
literally a single detected photon per pixel~\cite{FPI2014} or about 1.0 detected photon per pixel on average~\cite{Shin2015,Altmann2016,Shin2016camera}
by exploiting probabilistic models for individual photon detections
and regularization inspired by typical scene structure.
A key contribution of~\cite{FPI2014,Shin2015} is the use of photon-by-photon processing
that attempts to remove the detections that are likely due to background noise.
While this \emph{censoring} is also an exploitation of spatial structure,
it is
introduced primarily to remove
a nonconvexity inherent to ML estimation of depth in the presence of background noise.
Furthermore, it is applied only to depth estimation---not to reflectivity estimation.
The censoring in~\cite{Shin2016camera}
is also applied only to depth estimation and
is based on the depths in the entire field of view being sparse after appropriate discretization.
While these methods are effective
in low-light scenarios where histogramming techniques perform poorly,
the imaging accuracy degrades significantly in the presence of high background noise.
This is of particular importance for long-distance or low-power measurements in daylight,
when the rate of photon detection from ambient light may be significantly higher than the detection rate from the active illumination.

Building primarily upon~\cite{Shin2015}, this paper reexamines the model of low-flux detection as an inhomogeneous Poisson mixture process.
Given that estimation from few detections has been demonstrated when signal and noise levels are equal, we aim to use new insights from the model to make accurate imaging possible when noise levels are 25 times higher than signal
(with other imaging conditions unchanged).
The central idea is that by effectively separating the signal and noise contributions, estimates can be computed that are almost as good as an oracle that uses only the signal detections.
Here we focus on using detection times and intuition from the Poisson process model to approximately unmix signal and noise contributions at each pixel.
We also introduce spatial adaptivity to overcome low-reliability depth estimates when too few signal photons are detected.
While some key concepts are first introduced within a pixelwise ML estimation framework,
as in previous works~\cite{FPI2014,Shin2015,Shin2016multidepth,Shin2016camera},
we ultimately apply regularization to improve image formation.
In Section~\ref{sec:setup}, we give a brief overview of the experimental setup, the probabilistic model of photon detection, and the ML estimators for reflectivity and depth.
Section~\ref{sec:unmix} motivates the use of windowing for imaging at low signal-to-background ratio\footnote{SBR
is the mean detections due to backreflected light divided by the mean detections due to ambient light and dark counts.
It is used rather than signal-to-noise ratio to be clear that Poisson variations in numbers of detected signal are not considered \emph{noise}.} (SBR) by limiting the false acceptance of background detections as signal detections.
It also motivates a spatially-adaptive approach and discusses how one can fill in information for pixels with too few signal detections.
Section~\ref{sec:algorithm} introduces our unmixing algorithm, built into the framework of~\cite{Shin2015}.
We demonstrate the algorithm's performance on both simulated and real data in Section~\ref{sec:results}.
Finally, Section~\ref{sec:conclusion} presents our conclusions and suggestions for further work.

\section{Data Acquisition, Modeling, and Baseline Estimators}\label{sec:setup}
The methods of this paper are applicable to a generic raster-scanning LIDAR system.
Dwell times are fixed, so acquisition could be parallelized with simultaneous
illumination of multiple scene patches and array detection.%
\footnote{This could be called \emph{pseudoarray} imaging in that it idealizes the array rather than to compensate for array-specific non-uniformities~\cite{Shin2016camera}.
}
We review specifically the experimental data collection, modeling,
and image formation methods of~\cite{Shin2015},
as it is most closely related to the present work and
the publicly distributed data associated with that paper will be used to validate our new methods.
The scenes used in~\cite{Shin2015} are rather simple.
To validate our methods on more complicated scenes, we also
simulate the data collection method of~\cite{Shin2015} on scenes from~\cite{scharstein2007}.

\subsection{Experimental Setup}
\begin{figure}
\centering
\includegraphics[width=0.9\linewidth]{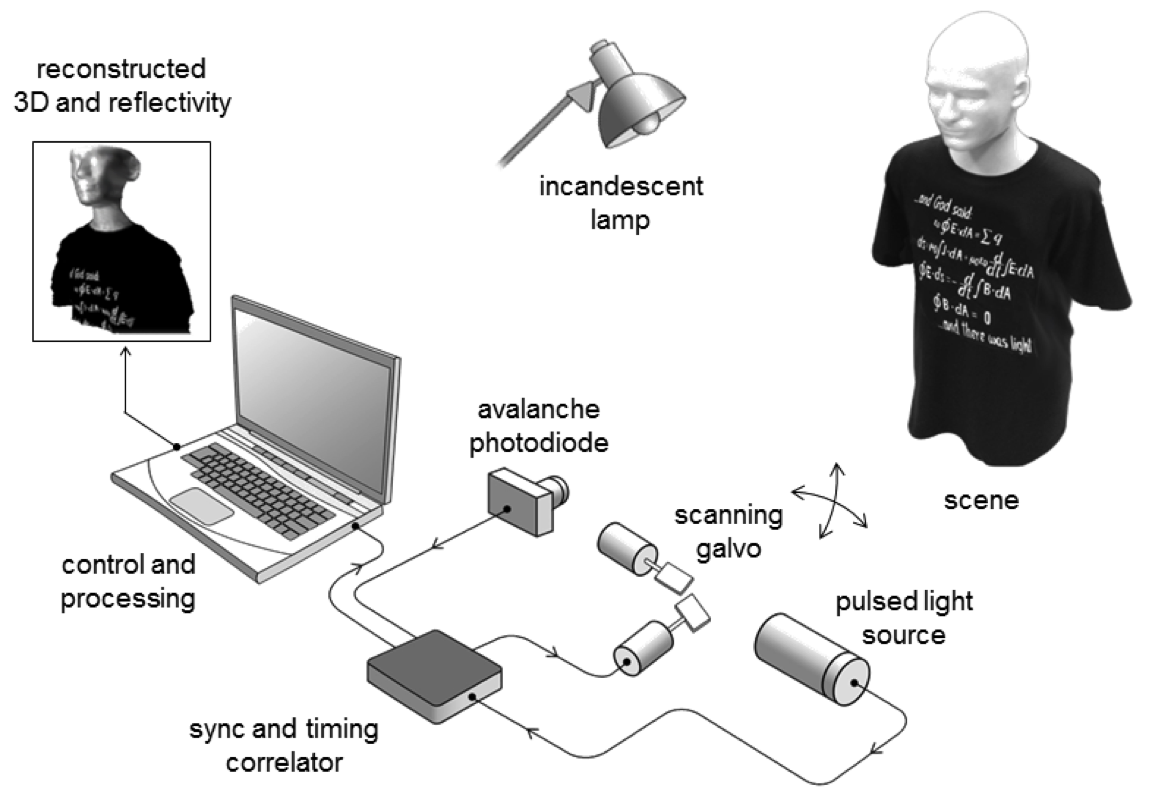}
\caption{Experimental imaging setup for photon-counting LIDAR\@. Photon detections correlated with laser pulse times yield time-of-flight data, and the number of photon returns indicate scene reflectivity.  Reproduced from~\cite[Fig.~2]{Shin2015}.}
\label{fig:setup}
\end{figure}

The setup in Figure~\ref{fig:setup} was used to raster-scan a scene by directing a pulsed laser at each patch $(i,j)$ in a scene via a two-axis galvanometer.
The illumination has a pulse shape $s(t)$ with RMS duration $\Tp$.
Each scene patch corresponds to one pixel in our depth image $\bz \in \R^{N_i\times N_j}_+$ and reflectivity image $\balpha \in \R^{N_i\times N_j}_+$.
The reflectivity $\alpha_{i,j}$ includes the effects of radial fall-off, view angle, and material properties of patch $(i,j)$.
Each patch is illuminated with $\Nr$ pulses at a repetition period of $\Tr$.
To prevent distance aliasing, we ensure that
$\Tr > {2 \zmax}/{c}$, where $\zmax$ is the maximum scene depth and $c$ is the speed of light.
The SPAD detector triggers a picosecond-resolution time stamp when a photon is detected, marking the time $t_{i,j}$ relative to the previous laser pulse. 
The full vector of photon detections at $(i,j)$ is given as $\{t_{i,j}^{(\ell)}\}_{\ell=1}^{k_{i,j}}$,
where $k_{i,j}$ is the total number of photons detected at that pixel. 

Detector counts may be the result of laser-pulse photons back-reflected from the scene,
ambient photons emitted by an incandescent lamp, or SPAD dark counts not caused by incident photons.
After a detection, a SPAD detector has a \emph{reset time} or \emph{dead time} during
which there is no sensitivity to incident light.
Because of dead time, at most one detection event
is recorded for each pulse-repetition period.

Further acquisition details can be found in~\cite{FPI2014} and its supplement~\cite{FPI2014supp},
with the modifications for the use of fixed dwell time in~\cite{Shin2015}.

\subsection{Probabilistic Measurement Model}\label{sec:measurement}
As described in~\cite{Shin2015}, the illumination of pixel $(i,j)$ with a pulse $s(t)$ results in photon flux at the detector described by
\begin{equation}
    r_{i,j}(t) = \alpha_{i,j}s(t-2z_{i,j}/c) + b_\nu,
\end{equation}
where $b_\nu$ is the flux due to ambient light at the optical operating frequency $\nu$.
At the detector, this photon flux is reduced by the detector's quantum efficiency $\eta \in [0,1)$, which describes the probability that an incident photon is registered by the device.
Detector dark counts are added at rate $\bd$, resulting in a total detection \emph{intensity} given by
\begin{subequations}
\begin{align}
    \lambda_{i,j}(t) &= \eta r_{i,j}(t)+\bd \\
        &= \eta \alpha_{i,j}s(t-2z_{i,j}/c) + (\eta b_\nu + \bd),
\end{align}
\end{subequations}
which groups the non-informational noise terms together and ignores any effect of detector dead time.

Over one illumination period, the detection \emph{rate} is 
\begin{subequations}
\begin{align}
    \Lambda(\alpha_{i,j}) &= \int_0^{\Tr} \lambda_{i,j}(t) \, dt \\
        &= \eta \alpha_{i,j}S+B,
\end{align}
\end{subequations}
where we define $S = \int_0^{\Tr} s(t) \, dt$ and $B = (\eta b_\nu + \bd)\Tr$.

Operating in a low-flux regime, we have $\eta \alpha_{i,j}S+B \ll 1$, so the probability of a detection in any given illumination period is small, and the probability of multiple detections in one period is negligible.%
\footnote{Low-flux operation is a requirement of time-correlated single photon counting (TCSPC) systems due to detector and electronics dead times. To avoid missed detections and a bias towards early detections times, the manufacturer of the TCSPC system suggests average count rates should be limited to at most 1\% to 5\% of the illumination periods~\cite{tcspcWahl2015}.
}
Thus, by restricting our operation to the low-flux regime, we incur little error by ignoring the detector dead time effects throughout the model.
In particular, by ignoring dead time, the detections are an inhomogeneous Poisson process with intensity $\lambda_{i,j}(t)$~\cite{Snyder1991}.
As a result, each detection time is an independent, identically distributed random variable $U$ with common probability density
\begin{equation}
    p_U(u) = \frac{\lambda_{i,j}(u)}{\Lambda(\alpha_{i,j})}, \qquad u \in [0,\Tr).
\end{equation}

Unlike in~\cite{FPI2014},
we observe the detection process for a fixed, deterministic number of illumination repetition periods;
thus, no information is conveyed by the order of the detection times or the identities of the repetition intervals in which the detections occur.
It is convenient to exploit the periodicity of $\lambda_{i,j}(t)$ to fold time interval
$[0, \Nr \Tr)$ down to $[0, \Tr)$ to obtain an equivalent model in which all detections occur within one illumination period due to a process with intensity
\begin{equation}\label{eq:nr_intensity}
    \lambda_{i,j}^{\Nr}(t)=\Nr[\eta \alpha_{i,j}s(t-2z_{i,j}/c) + (\eta b_\nu + \bd)]
\end{equation}
and rate
\begin{equation}
    \Lambda_{\Nr}(\alpha_{i,j}) = \Nr (\eta \alpha_{i,j}S+B).
\end{equation}
The distribution of photon counts is
\begin{equation}\label{eq:K_Poisson}
    K_{i,j} \sim \Poisson (\Lambda_{\Nr}(\alpha_{i,j}))
\end{equation}
and the probability density of detection times is
\begin{equation}\label{eq:dt_pdf}
    p_{T_{i,j}}(t) =  \frac{\lambda^{\Nr}_{i,j}(t)}{\Lambda_{\Nr}(\alpha_{i,j})},
    \qquad t \in [0,\Tr).
\end{equation}

It is useful to decompose $\lambda_{i,j}^{\Nr}(t)$ into two independent processes.
A \emph{signal process} is inhomogeneous with intensity
\begin{equation}
\label{eq:lambda-s}
\lambdas(t) = \Nr \eta \alpha_{i,j}s(t-2z_{i,j}/c),
\end{equation}
and a \emph{background process} is homogenous with intensity
\begin{equation}
\label{eq:lambda-b}
\lambdab(t) = \Nr(\eta b_\nu + \bd).
\end{equation}
At each $(i,j)$, the number of detections due to signal is 
\begin{equation}\label{eq:poiss_noise}
     M_{i,j} \sim \Poisson (\Nr \eta \alpha_{i,j} S),
\end{equation}
and the number of detections due to noise is 
\begin{equation}
     N_{i,j} \sim \Poisson (\Nr B).
\end{equation}

\subsection{Binomial vs.\ Poisson Modeling}
The duration of a typical SPAD detector dead time is similar to a typical repetition period $\Tr$.
Thus, as noted earlier, at most one detection event is recorded for each pulse-repetition period.
As developed in~\cite{Shin2015},
under the simplifying approximation that a dead period ends at the subsequent pulse-repetition boundary,
this makes detection within each pulse-repetition period a Bernoulli trial and the total number of detections in $\Nr$ pulse-repetition periods a binomial random variable.
More precisely,
\begin{subequations}\label{eqs:binomial_model}
\begin{equation}\label{eq:K_binomial}
     K_{i,j} \sim \binomial(\Nr,1-P_0),
\end{equation}
where
\begin{equation}
    P_0(\alpha_{i,j}) = \exp[-(\eta \alpha_{i,j} S +B)]
\end{equation}
\end{subequations}
is the probability of zero detections in one pulse-repetition period.

Under a low-flux assumption, the models \eqref{eq:K_Poisson} and \eqref{eqs:binomial_model} for $K_{i,j}$ are approximately equal;
a formal equivalence can be shown through the Poisson limit theorem.
The binomial model is perhaps slightly more accurate because it ignores only the portion of the dead period that falls after the subsequent pulse-repetition boundary,
but this is negligible by assumption and introduces a bias
(which is again negligible by assumption).

A possible downside of the binomial model is philosophical:
it encourages one to discard the detection times when estimating reflectivity,
as is done in~\cite{FPI2014,Shin2015,Shin2016camera}.
The Poisson model instead encourages the separation into signal and background processes,
which leads to a separation of $K_{i,j}$ into its constituents $(M_{i,j},N_{i,j})$;
estimation of reflectivity from $M_{i,j}$ is more accurate than from $K_{i,j}$, especially when SBR is low.

\subsection{Parameter Estimation}
\subsubsection{Binomial Model of Detection}

The binomial model \eqref{eqs:binomial_model} results in a constrained ML (CML) reflectivity estimate given by
\begin{equation}\label{eq:shin_alpha_ml}
    \hat{\alpha}_{i,j}^{\rm CML} = \max \left \{\frac{1}{\eta S}\left[\log \left ( \frac{\Nr}{\Nr-k_{i,j}} \right ) -B \right ],0\right \}.
\end{equation}
This expression essentially counts the number of detections $k_{i,j}$ out of the $\Nr$ illumination intervals and subtracts a constant offset $B$, the expected number of noise detections per illumination period.
For $B$ characterized as either a Poisson or binomial random variable, it is a low-variance estimator of the true number of noise detections $n_{i,j}$ as long as $B$ is small, since the mean and variance of these random variables are proportional.
The problem with this estimator arises when $\eta \alpha_{i,j} S$ remains small but $B$ is significantly larger. 
In that case, the variance of $B$ is also increased, so its reliability as an estimator of $n_{i,j}$ decreases.

\subsubsection{Reflectivity Estimation from a Poisson Process}
By instead approaching detection entirely as an inhomogeneous Poisson process, we can take advantage of the detection times in both the reflectivity and depth estimates.
The signal back-reflected from $(i,j)$ is approximately the illumination pulse with intensity modulated by the reflectivity $\alpha_{i,j}$.
Estimation of $\alpha_{i,j}$ then requires the same approach as estimation in amplitude-modulated optical communication as described in~\cite{BarDavid1969}.

The likelihood function for the set of observed photon detections $\{t_{i,j}^{(\ell)}\}_{\ell = 1}^{k_{i,j}}$ is
\begin{equation*}
    p\left[\{t_{i,j}^{(\ell)}\}_{\ell = 1}^{k_{i,j}} \,;\, \alpha_{i,j},z_{i,j}\right]
    = e^{-\Lambda_{\Nr}(\alpha_{i,j})} \prod_{\ell=1}^{k_{i,j}}  \lambda_{i,j}^{\Nr}(t_{i,j}^{(\ell)}),
\end{equation*}
which yields a CML estimate given by 
\begin{align}
    \hat{\alpha}_{i,j}^{\rm CML} = \underset{\alpha_{i,j}\geq 0}{\operatorname{arg\,max}}  \sum_{\ell = 1}^{k_{i,j}} \log &\big[ \Nr(\eta \alpha_{i,j} s(t_{i,j}^{(\ell)}-2z_{i,j}/c) \nonumber \\ 
    & + (\eta b_\nu + \bd)) \big] - \Lambda_{\Nr}(\alpha_{i,j}).
\end{align}
Differentiating with respect to $\alpha_{i,j}$, we have
\begin{equation}\label{eq:alpha_est}
        \sum_{\ell = 1}^{k_{i,j}} \frac{\eta s(t_{i,j}^{(\ell)}-2z_{i,j}/c)}{\eta \alpha_{i,j} s(t_{i,j}^{(\ell)}-2z_{i,j}/c) + (\eta b_\nu + \bd)} = {\Nr}\eta S.
\end{equation}
Since all terms in \eqref{eq:alpha_est} are nonnegative, the left-hand side is monotonically decreasing in $\alpha$, so a unique optimal estimate of $\alpha_{i,j}$ exists.
Unfortunately, this expression requires knowledge of the true depth for the optimal estimate.

At high SBR, an approximate solution is given by 
\begin{equation}\label{eq:alpha_est2}
    \hat{\alpha}_{i,j}^{\rm CML} = \max \left \{ \frac{k_{i,j}-\Nr B}{\Nr \eta S},0 \right \},
\end{equation}
which preserves the non-negativity of $\alpha$ and simplifies to 
\begin{equation}
    \hat{\alpha}_{i,j}^{\rm ML,noise-free} = \frac{k_{i,j}}{{\Nr}\eta S}
\end{equation}
if noise is completely eliminated.
Conveniently, these estimates have a closed form solution, which is simply the normalized photon count.
When noise is low, all detections are due to signal, so the count is a sufficient statistic for the reflectivity---we no longer need to use the detection times or know the true depth $z_{i,j}$.

\subsubsection{Depth Estimation}
The process of depth estimation from the set of detection times is also derived in~\cite{BarDavid1969}. 
The CML depth estimate is given by
\begin{align} \label{eq:z_est}
    \hat{z}_{i,j}^{\rm CML} = \argmax_{z_{i,j}\in[0,\zmax)} \sum_{\ell=1}^{k_{i,j}}\log\big[\eta &\alpha_{i,j} s(t_{i,j}^{(\ell)}-2z_{i,j}/c ) \nonumber \\ 
    &+(\eta b_\nu + \bd) \big].
\end{align}
We can see that this requires knowledge of the true $\alpha_{i,j}$ value, and furthermore that the noise term adds a nonconvexity.
In practice, $\hat{z}_{i,j}$ is computed by finding the delay that maximizes the output of a log-matched filter. 

Again we remark on the case of zero noise, where the depth estimate is given by
\begin{equation}
    \hat{z}_{i,j}^{\rm CML} = \argmax_{z_{i,j}\in[0,\zmax)} \sum_{\ell=1}^{k_{i,j}}\log\left[s(t_{i,j}^{(\ell)}-2z_{i,j}/c ) \right].
\end{equation}
In this case the noise-free solution is also greatly simplified, as it is convex and has no dependence on $\alpha_{i,j}$.

\section{Unmixing Signal and Noise Processes}\label{sec:unmix}
\begin{figure*}
    \centering
    {\phantomsubcaption\label{fig:proc_pdf}}
    {\phantomsubcaption\label{fig:sample_dets}}
    {\phantomsubcaption\label{fig:sample_spxl}}
    {\phantomsubcaption\label{fig:sample_form_spxl}}
    \includegraphics[trim={0 5.5cm 0 1.2cm},clip, width=0.9\linewidth]{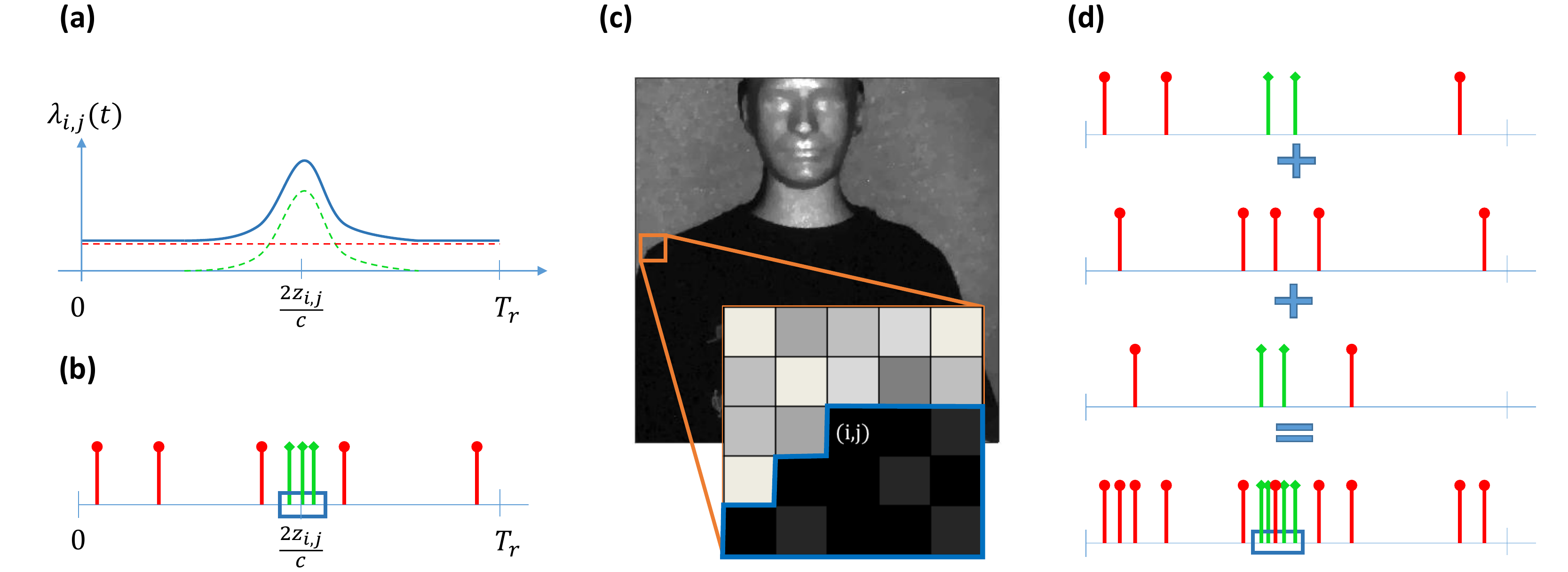}
    \caption{\subref{fig:proc_pdf} Detection can be described as an inhomogeneous Poisson process (shown in blue), which is the sum of  inhomogeneous signal (green) and homogeneous background (red) processes. As in the example set of detections in \subref{fig:sample_dets}, signal detections tend to form clusters more readily than the background detections, suggesting windowing as an approach to unmixing signal and noise. Pixels with similar transverse position and reflectivity tend to belong to the same object and therefore have similar depth, as in \subref{fig:sample_spxl}. Using this observation leads to borrowing detections from similar neighboring pixels, as illustrated in \subref{fig:sample_form_spxl}, which can help amplify low signal levels by making signal detection clusters more apparent.}
    \label{fig: }
\end{figure*}

The key observation from the parameter estimates is that the reflectivity and depth estimates are coupled and complicated in the presence of noise, but both are greatly simplified if noise is removed. 
Indeed, if we could unmix detection into its component signal and background processes, we could ignore the noise detections and simply apply the noise-free estimators.
Rather than a conventional approach of forming estimates first and then denoising, this observation is motivation for separating signal from noise first and then forming estimates.

\subsection{Pixelwise Unmixing}
At an individual pixel, no marker distinguishes between signal and noise detections, so no explicit information is available to separate the signal from the noise.
In order to unmix the processes, the only information we have \emph{a priori} is the different probabilistic models of the detection processes.
The signal process rate is related to the short-duration illumination pulse, so signal detection times have a small variance.%
\footnote{Technically, we are referring to a small variance for any fixed value of the true depth $z_{i,j}$.
In a Bayesian formulation in which $z_{i,j}$ has a prior distribution,
the signal detection times have a small conditional variance given $z_{i,j}$ or with other conditioning that approximately localizes $z_{i,j}$.}
This suggests that when several signal photons are detected at the same pixel, the detections will be clustered together near the true depth, as illustrated in Figures~\ref{fig:proc_pdf} and~\ref{fig:sample_dets}.
The background process has a constant rate, meaning no time is more likely than any other to have a background detection.
Since the background photons are uniformly distributed in time, we expect them to be fairly spread out in general, unless the background detection rate is very high.

Since signal detections tend to cluster together more readily than background detections, an intuitive approach to identifying signal photons is to search for the largest of those clusters.
One way to define a cluster of detections is to choose a window of duration $\Twind$ and a minimum cluster size $\Ncl$.
The window duration should be chosen such that $\Tp < \Twind \ll \Tr$, so that a well-placed window
(one shifted by approximately $2z_{i,j}/c$) is large enough to capture most or all signal detections, without accepting too many noise detections.
If at $(i,j)$ there are at least $\Ncl$ detections within some window of duration $\Twind$, then we can consider $(i,j)$ as having a cluster of detections. 
If there happen to be multiple clusters at $(i,j)$, we choose the window with the most detections $\Nmax$ as our signal cluster.
From the shift of the window, we have an estimate of the depth $z_{i,j}$,
and since the mean number of noise detections in a short interval $\Twind$ is close to zero,
$\Nmax$ yields a rather accurate estimate of the number of signal detections $m_{i,j}$
analogously to \eqref{eq:alpha_est2}.
As detailed later, the purpose of the $\Ncl$ minimum---rather than to seek the largest cluster regardless of its size---is to have a mechanism to produce no depth estimate rather than an unreliable one.

In fact, this intuitive windowing approach falls out of the ML reflectivity and depth estimates.
Crudely approximating $s(t)$ by a square wave of duration $\Tp$ centered at $2z_{i,j}/c$, the reflectivity estimate in \eqref{eq:alpha_est} is due only to detections that occur within $\Tp/2$ of the true depth. 
Even for a more realistic pulse shape, only detections within a short duration around the true depth contribute non-negligible weight to the reflectivity estimate.
Furthermore for the depth estimate, again approximating $s(t)$ as a square wave, the log-matched filter is maximized at the window containing the largest number of detections.

\begin{figure*}
    \centering
    \begin{subfigure}{0.32\linewidth}
        \centering
        \includegraphics[trim={2mm 0 2mm 7mm},clip, width=\textwidth]{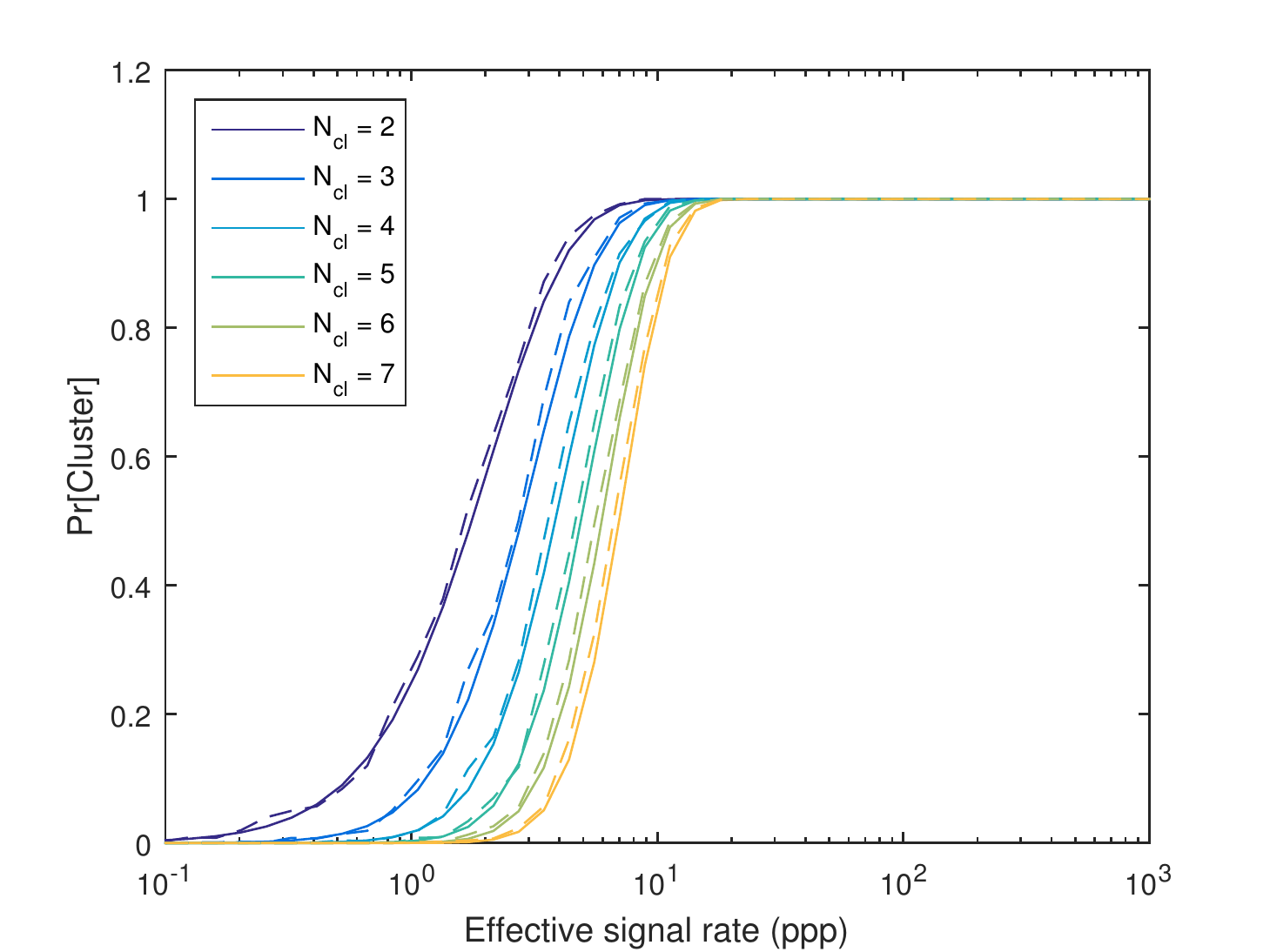}
        \caption{}
        \label{fig:sig_clust}
    \end{subfigure}
    \begin{subfigure}{0.32\linewidth}
        \centering
        \includegraphics[trim={2mm 0 2mm 7mm},clip, width=\textwidth]{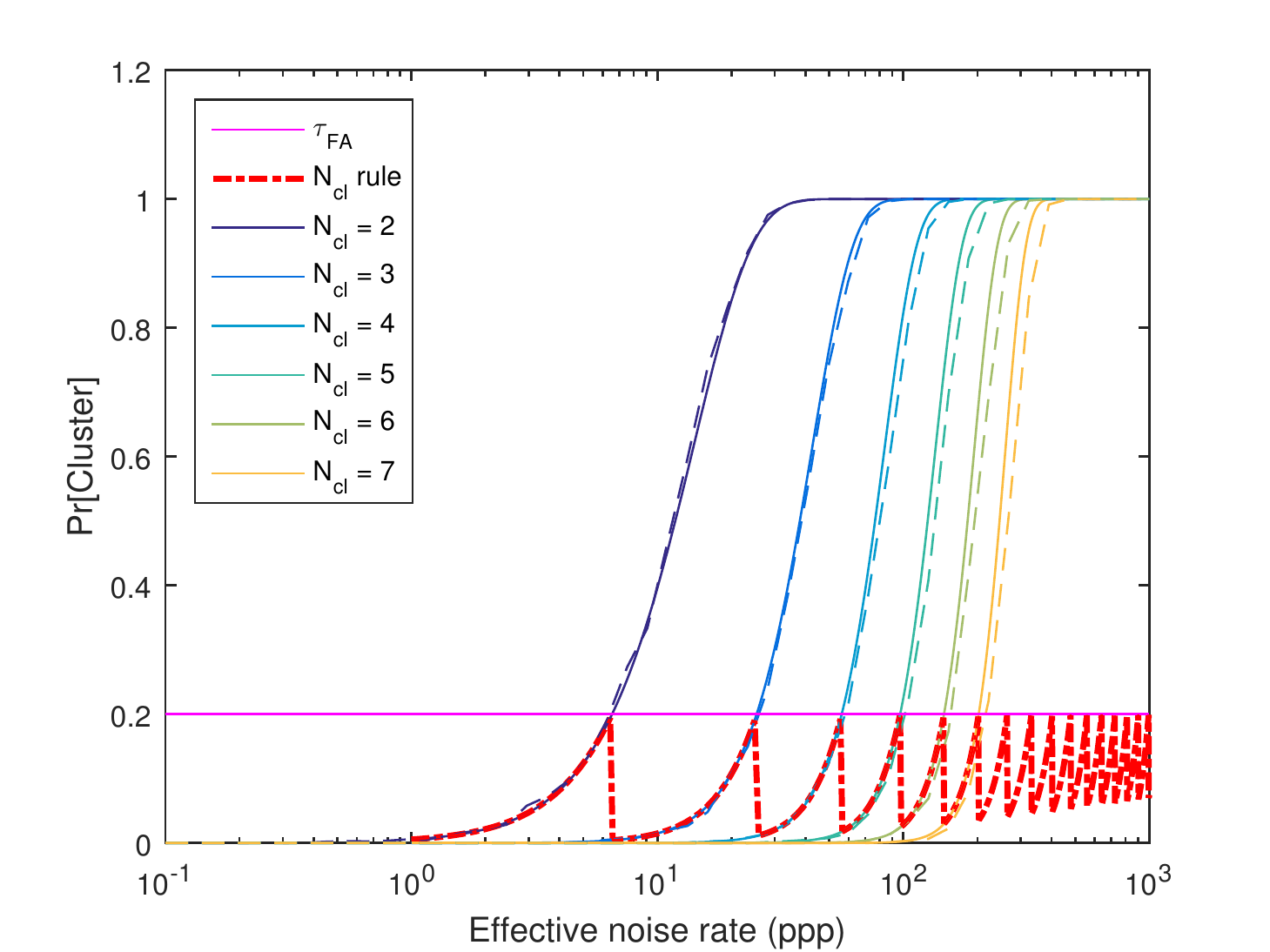}
        \caption{}
    \label{fig:noise_clust}
    \end{subfigure}
    \begin{subfigure}{0.32\linewidth}
        \centering
        \includegraphics[trim={2mm 0 2mm 7mm},clip, width=\textwidth]{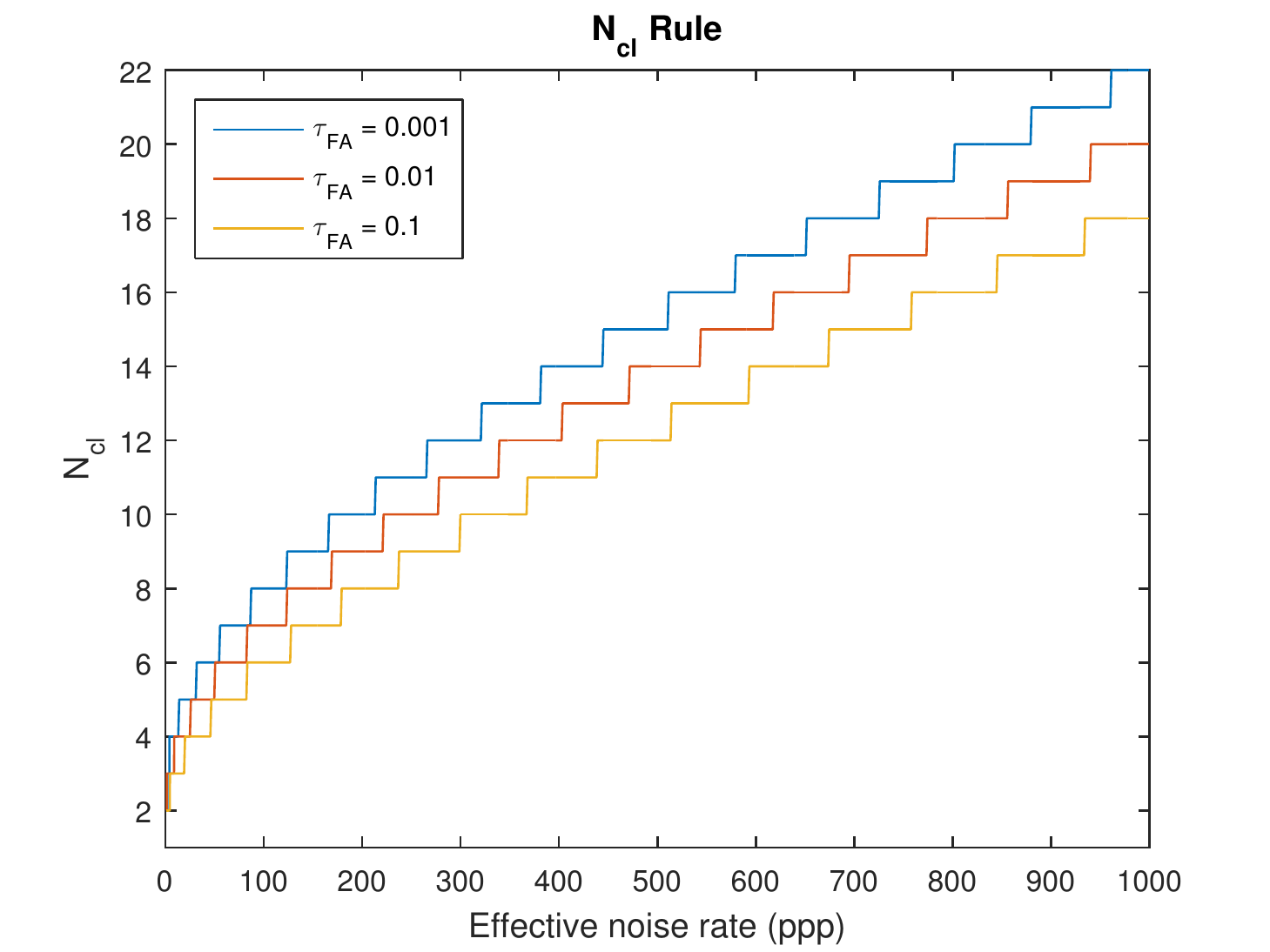}
        \caption{}
        \label{fig:ncl_rule}
    \end{subfigure}
    \caption{In comparing the theoretical approximations (solid lines) and Monte Carlo simulation (dashed lines) of the probability of cluster occurrence due to signal (a) or noise (b) at various values of $\Ncl$, it is apparent that the derivations in
    the appendix
    give close approximations to the true clustering probabilities.
    Since only the effective noise rate is known, setting a performance threshold $\tauFA$ as in (b) yields a cluster size rule in (c) that limits acceptance of noise clusters to probabilities less than or equal to $\tauFA$.}
    \label{fig:clust_prob}
\end{figure*}

In the appendix, we derive the separate probabilities that the signal and background processes will generate detection clusters, according to our definition. 
Figures~\ref{fig:sig_clust} and~\ref{fig:noise_clust} compare these derivations to Monte Carlo simulations of clustering based on the detection model, confirming that these derivations produce reasonable probability estimates and that the simplifying assumptions are minor. 
For all experiments and derivations, $\Twind$ was fixed to $2\Tp$, where $\Tp = 270$~ps is the measured RMS pulse width of the experiments in~\cite{FPI2014}.
This window size covers more then 95$\%$ of the probability mass of signal detection for a Gaussian pulse shape approximation.
The pulse repetition period $\Tr = 100$~ns is also used from~\cite{FPI2014}.

Using the plots of these probabilities in Figure~\ref{fig:clust_prob},
we observe that for some rates of signal and background detection, our intuition of finding clusters of detections by windowing is justified.
For instance, if the signal and background rates were each 10 photons per pixel (ppp),
the probability of observing a cluster of signal detections would be about 1 for any of the minimum cluster sizes shown (see Figure~\ref{fig:sig_clust}),
whereas the probability of observing a cluster of background detections would be negligible for $\Ncl > 2$ (see Figure~\ref{fig:noise_clust}). 
As a result, the largest cluster of detections is likely to have more than two detections, and thus the largest cluster could be safely assumed to contain at least one signal detection.

Ideally, we could select optimal $\Twind$ and $\Ncl$ values for each pixel based on the local signal and noise rates. 
However, we only know the background rate from calibration; the signal rate is unknown since it depends on the unknown reflectivity parameter.
Our approach from the given data is to fix a reasonable window size and choose $\Ncl$ based solely on the noise rate, which is known from calibration testing with no signal input.
Given that we assume the largest cluster at a pixel is due to signal, we restrict our minimum cluster size to limit the number of clusters falsely accepted as signal when they are actually due to noise.
As in Figure~\ref{fig:noise_clust}, we set a threshold $\tauFA$ for the probability of clusters due to noise that we will allow. 
For any given noise rate, we can then choose the smallest $\Ncl$ that will yield $\Pr[\mbox{noise cluster}] < \tauFA$.
This method of choosing $\Ncl$ as a function of the noise rate is illustrated in Figure~\ref{fig:ncl_rule}.
Since the theoretical derivation tends to slightly overestimate $\Pr[\mbox{noise cluster}]$, we are likely to see even fewer clusters due to noise than the actual threshold we set.

Now that a reasonable cluster definition is established, we can window the detections at each pixel, and if $\Nmax>\Ncl(\Nr B,\tauFA,\Twind)$, we discard all detections except those in the best window as noise.

\subsection{Spatially-Adaptive Unmixing}
While setting a low $\tauFA$ is a good approach for limiting the number of accepted noise clusters, the resulting $\Ncl$ may be too high for any cluster to be found.
Moreover, for photon-efficient imaging, it is common for large regions of scenes to have very few signal detections, such as in \cite{Shin2015}, where some scenes were reported as having $54\%$ of pixels with no detections.
Even with no background present, it would be impossible to estimate the depth purely from windowing, since there are no clusters to identify.
As a result, relying on windowing for low-$\alpha$, low-SBR data often yields too many pixels with no depth estimate.

A key insight into solving this problem comes from analyzing the behavior of ML depth estimates in noise.
\begin{figure}
    \centering
    \includegraphics[width=\linewidth]{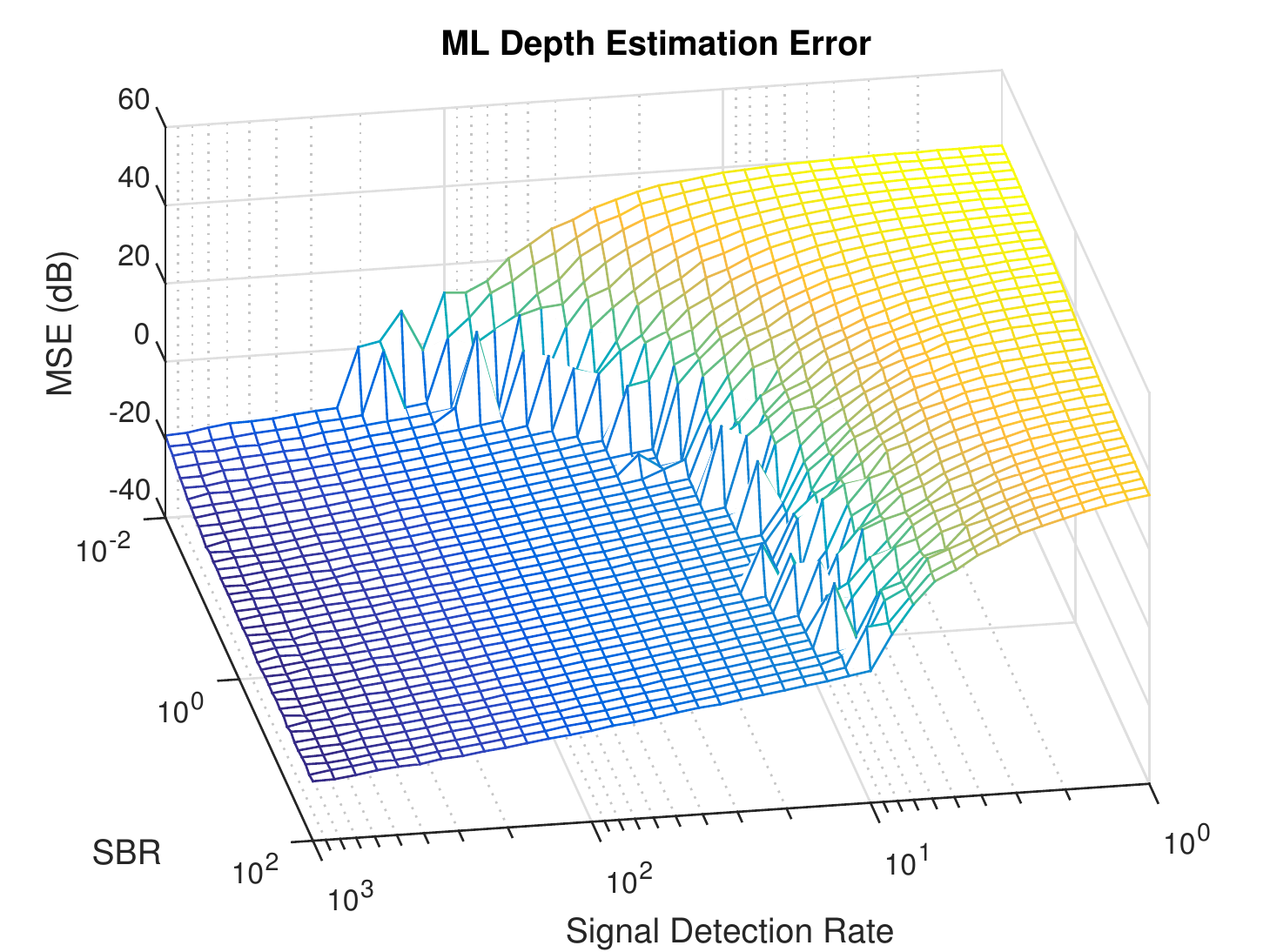}
    \caption{Results of a Monte Carlo simulation of ML depth estimates where the noise is uniformly distributed over the acquisition interval and $s(t)$ is approximated by a Gaussian pulse of RMS width approximately equal to our measured laser pulse.}
    \label{fig:monte_carlo}
\end{figure}
Figure~\ref{fig:monte_carlo} shows the results of Monte Carlo simulations of depth estimation for various $\alpha$ and SBR values, where we see a thresholding behavior that is common to nonlinear estimators \cite{Erkmen2009}.
Specifically, if enough signal detections are present, the ML depth estimate has low mean-squared error, regardless of the SBR\@. 
This phenomenon is due to the small variance of the signal process relative to the noise, resulting in a strong peak where large numbers of signal detections cluster together, even when $m_{i,j} \ll n_{i,j}$.
The natural solution is that used by conventional LIDAR: by acquiring many detections over long acquisitions times, the coherent signal becomes easier to detect in incoherent noise, even if the signal is relatively weak.

Given the purpose of photon-efficient imaging, extending acquisition time is an inadequate solution, but a few further observations allow this method to be used as inspiration.
First, natural scenes, especially depth maps, are generally smooth except at object boundaries, so neighboring pixels often have approximately equal depth.
This was the justification for the total variation (TV) regularization used in~\cite{Shin2015}, since TV regularization tends to smooth out noise while still preserving jump discontinuities~\cite{louchet2008}. 
Secondly, edges in reflectivity and object boundaries in depth tend to be co-located, so scene patches that are similar in both reflectivity and transverse position likely are similar in longitudinal position (depth) as well.

These observations can be codified through the construction of \emph{superpixels}, oversegmentations of an image into small regions of similar pixels, which are a common tool in computer vision applications.
Superpixels were originally introduced in \cite{ren2003learning} with the idea that pixels are arbitrary elementary units of digital images, and that breaking images into more natural building blocks could improve and speed up further processing such as larger-scale image segmentation and object detection.
The general reasoning is that pixels that are similar in both some color space (e.g., Lab) and in transverse position have a high probability of belonging to the same object. 
As a result, superpixels have been used as a preprocessing tool to provide noise robustness and to fill in gaps of depth maps for stereo \cite{zitnick2007stereo,cambra2014improving}, RGB-d \cite{van2013depth}, and LIDAR \cite{Mahmoudabadi} systems.

Our approach is to use a variant of superpixels to artificially extend acquisition times, which facilitates depth estimation.
Consider a small neighborhood of pixels in a scene, such as the one illustrated in Figure~\ref{fig:sample_spxl}.
Assuming that the scene has been sampled with adequate transverse spatial resolution, pixels within this neighborhood will have similar depth values, unless the neighborhood crosses a boundary between objects.
If $(x,y)$ is in the neighborhood of $(i,j)$, then
$p_{T_{i,j}}(t) \approx p_{T_{x,y}}(t)$ for all $t \in [0,\Tr)$.
Thus, combining detections from $(x,y)$ into the $(i,j)$ vector is almost equivalent to doubling the acquisition time at $(i,j)$.
This borrowing will maintain SBR but increase $m_{i,j}$, helping to reduce the estimation error, as we observed in Figure~\ref{fig:monte_carlo}.
Borrowing creates some smoothing in the transverse directions, and the ideal trade-off between noise reduction and this smoothing probably occurs just to the left of the estimation threshold illustrated in Figure~\ref{fig:monte_carlo}.

For practical purposes, reinforcing the coherent signal by borrowing detections from neighboring pixels will enhance the size of signal clusters and make windowing more reliable and useful, as illustrated in Figure~\ref{fig:sample_form_spxl}.
When superpixels are formed, the noise rate is effectively amplified by $\Nsp$, the number of pixels that contributed to the enhanced detection vector at $(i,j)$, so we update our cluster size requirement for windowing to $\Ncl(\Nsp\Nr B,\tauFA,\Twind)$ to avoid falsely accepting noise clusters.
This formulation in fact describes the generic windowing procedure, where $\Nsp =1$ if the detections at only a single pixel are used.

There are many existing superpixel definitions and implementations, each designed to meet particular performance criteria \cite{achanta2012slic}.
In principle, any definition could be used within our algorithm to select groups of similar pixels from which to borrow photon detections.
A key difference of our approach to that of the other depth-estimation applications of superpixels in \cite{zitnick2007stereo,cambra2014improving,van2013depth,Mahmoudabadi} is that the existing methods all incorporate a conventional digital camera.
Superpixels are formed using the high-quality color images, and then the assumed redundancy of these regions is applied to corresponding regions of a lower-quality depth map to fill in gaps or filter out noise.
In our system, reflectivity is a grayscale value estimated from the same active illumination data used to estimate the depth. 
Due to the low signal counts and high background levels, the intensity data is much less reliable than that from a conventional camera.
As a result, we use a simple definition of selecting the subset of pixels in a square region that meet a reflectivity tolerance compared to $(i,j)$.
Since pixels are chosen within a fixed distance of $(i,j)$, the set of candidate pixels changes slightly from one pixel of interest to the next.
We use this particular definition in order to promote a high degree of localization, which helps preserve small changes in reflectivity and depth.
Other superpixel definitions that consider each region to be homogeneous would smooth over these small changes.

\section{Algorithm}\label{sec:algorithm}
\begin{figure}
\begin{subfigure}[c]{\linewidth}
    \includegraphics[trim={0 20pt 0 0},clip,width=\linewidth]{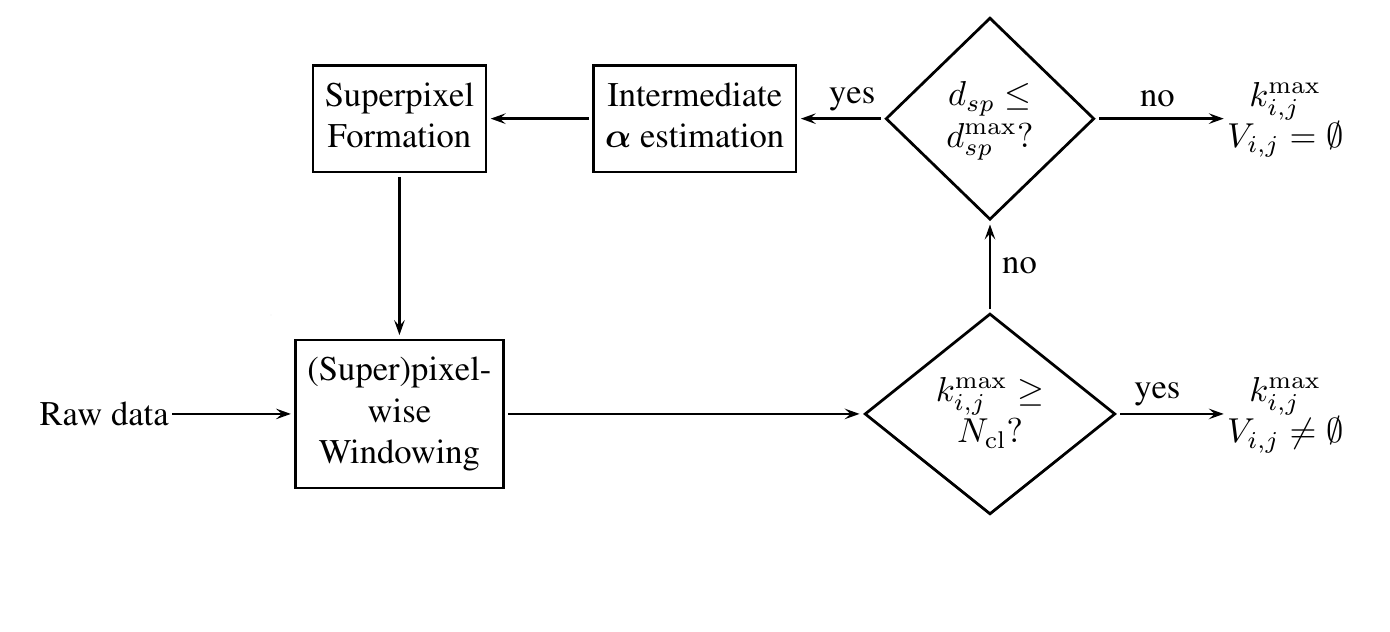}


\caption{Unmixing algorithm}
\end{subfigure}
\\
\begin{subfigure}[c]{\linewidth}
\centering %
\includegraphics[width=\linewidth]{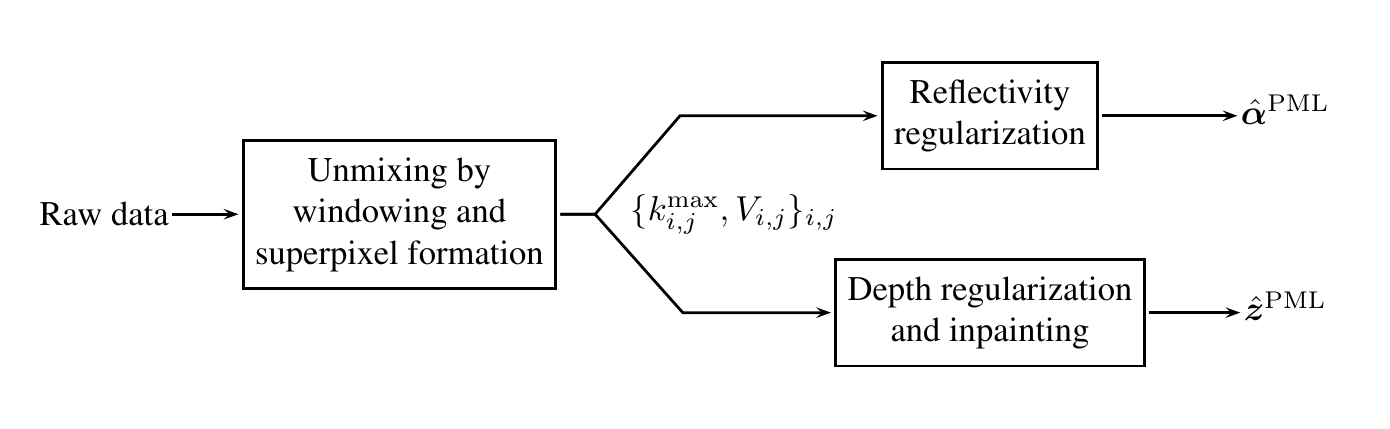}


\caption{Overall algorithm}
\end{subfigure}
\caption{Block diagram of the basic algorithmic sequence. The unmixing procedure continues until enough pixels have reliable depth estimates so that inpainting is needed for only a small fraction with missing entries.}
    \label{fig:block_diag}
\end{figure}

Our method for forming depth and reflectivity images from the raw detection data builds off the image formation procedure of \cite{Shin2015}, adding in the windowing and spatial adaptivity introduced in Section~\ref{sec:unmix}.
The procedure is summarized by the block diagram in Figure~\ref{fig:block_diag}, and we now detail each component.

The raw data input to the algorithm is the set of photon detections $\{t_{i,j}^{(\ell)}\}_{\ell=1}^{k_{i,j}}$ for each patch $(i,j)$ (thus implicitly including the values $k_{i,j}$).
It is also assumed that $B$,
the mean background count per patch per pulse-repetition period,
and $\eta S$,
the mean signal count per pulse in the absence of any attenuation,
have been measured through calibration or approximated from environmental conditions and hardware specifications.
A small number of algorithm parameters are introduced as needed.

\subsection{Windowing}\label{sec:windowing}
The process of 
noise censoring at patch $(i,j)$ by adaptive windowing uses two parameters:  a window length $\Twind$ and a target probability of false acceptance of a noise cluster $\tauFA$.  It is performed as follows, assuming for the moment no borrowing of detections from neighboring patches:
\begin{enumerate}

\item For each $\ell \in \{1,\dots,k_{i,j}\}$, find the set of detections in the interval of length $\Twind$ starting at the detection time $t_{i,j}^{(\ell)}$:
\begin{equation}
\mathcal{D}_\ell = \left\{t_{i,j}^{(k)}:t_{i,j}^{(\ell)} \leq t_{i,j}^{(k)} < t_{i,j}^{(\ell)}+\Twind\right\}.    
\end{equation}

\item Among these sets, select a set $W_{i,j}$ with the largest number of detections:
\begin{equation}
\mbox{$W_{i,j}$ satisfies $|W_{i,j}| = \displaystyle\max_{\ell} |\mathcal{D}_\ell|$},
\end{equation}
and define $\Nmax = |W_{i,j}|$.
(Resolve ties by choosing uniformly at random among the sets with $\Nmax$ detections.)

\item Using false acceptance threshold $\tauFA$, compute minimum cluster size $\Ncl$ as the smallest integer such that
\begin{equation}
    \Pr[\mbox{noise cluster}\,;\,\Ncl,\Nsp\Nr B,\Twind] < \tauFA,
\end{equation}
where $\Pr[\mbox{noise cluster}]$ is derived in \eqref{eq:noise_cluster} and $\Nsp = 1$ for windowing a single pixel.
Note that this step does not depend on the detection time data and thus desirable values of $\Ncl$ may be precomputed.

\item If $\Nmax\geq \Ncl$, retain only the detections that fall in the selected window $W_{i,j}$ and censor the rest, yielding the set of uncensored detections $\{t_{i,j}^{(\ell)}\}_{\ell \in V_{i,j}}$, where
\begin{equation}
V_{i,j} = \left \{\ell \in \{1,\dots,k_{i,j}\}: t_{i,j}^{(\ell)} \in W_{i,j} \right\}.
\end{equation}

\end{enumerate}
When this windowing is applied with superpixels (i.e., $\Nsp > 1$),
the detection time data $\{t_{i,j}^{(\ell)}\}_{\ell=1}^{k_{i,j}}$
is replaced by augmented detection times $\{\ttilde_{i,j}^{(\ell)}\}_{\ell=1}^{\ktilde_{i,j}}$.

\subsection{Reflectivity Estimation}\label{sec:refl_est}
In the window $W_{i,j}$, the expected number of noise detections is $\Nsp \Nr B \Twind/\Tr$, which is small even at low SBR and is considerably lower than the number of detections due to noise on the entire $[0,\Tr)$ interval.
Since noise detection is a homogeneous Poisson process, the variance in the number of noise detections in the window is also small, so $\Nsp \Nr B \Twind/\Tr$ is a good estimator of the number of noise detections.
Thus, we can modify \eqref{eq:alpha_est2} to estimate $\alpha_{i,j}$ from the window output as 
\begin{equation}
\hat{\alpha}_{i,j}^{\rm CML} = \max \left \{ \frac{\Nmax-\Nsp \Nr B \Twind/\Tr}{\Nsp \Nr \eta S},0 \right \}.
\end{equation}
For those pixels where $\Nmax < \Ncl$, this formula tends to slightly overestimate the reflectivity, since we have likely chosen the window with the largest cluster of noise detections.
However, the $\hat{\alpha}_{i,j}$ estimate is temporary for those pixels, since the value of $\Nmax$ will be updated after windowing the augmented data from the superpixels.

We form a reflectivity image by regularized ML estimation
with a regularization parameter $\beta_\alpha \in [0,\infty)$.
Using~\eqref{eq:K_Poisson}, the negative log-likelihood of the scene reflectivity $\alpha_{i,j}$ given the number of detections in $W_{i,j}$ is  
\begin{align}
\label{eq:L_alpha}
    \mathcal{L}_\alpha(\alpha_{i,j};\Nmax) = \Nsp \Nr \eta \alpha_{i,j} S \qquad \qquad \qquad \quad \qquad \nonumber \\
    - \Nmax \log \left[\Nsp(\Nr \eta \alpha_{i,j} S+\Nr B \Twind/\Tr) \right],
\end{align}
ignoring terms not dependent on $\alpha_{i,j}$.
As in~\cite{Shin2015}, we take advantage of spatial correlations in natural scenes to form a penalized ML (PML) estimate that enforces smoothness:
\begin{equation}
    \widehat{\balpha}^{\rm PML} = \argmin_{\balpha:\alpha_{i,j}\geq 0} \sum_{i=1}^{N_i} \sum_{j=1}^{N_j} \mathcal{L}_\alpha(\alpha_{i,j};\Nmax) + \beta_\alpha \pen_\alpha(\balpha).
\end{equation}

\subsection{Superpixel Formation}
After windowing, all pixels have a reflectivity estimate, but only those $(i,j)$ where $\Nmax \geq \Ncl$ have reliable depth estimates.
For those pixels with insufficient signal detection counts, superpixels are formed so that strongly correlated depth data from similar neighboring pixels can be combined to improve the performance of windowing. 
The key to our superpixel formation is to set bounds for what constitutes a similar pixel.
In this paper, superpixels borrow detections from all neighboring pixels within a fixed distance and a fixed reflectivity tolerance of our pixel of interest.
In particular, fix a neighborhood distance $\dsp$ (typically 1, 2, or 3) and a reflectivity tolerance $\tausp$ (typically around $5\%$ of the full range of $\hat{\balpha}^{\rm PML}$ values).
The superpixel at $(i,j)$ is defined as
\begin{align}
  \Neighborhood_{i,j} = \Big\{ & (x,y) \in \{1,\ldots,N_i\} \times \{1,\ldots,N_j\} \, :
  \nonumber \\
  & |i-x| \leq \dsp, \quad
    |j-y| \leq \dsp, \quad
  \nonumber \\
  & |\alphahatPML_{i,j}-\alphahatPML_{x,y}| \leq \tausp \Big\}.
\end{align}
The set of superpixel detections $\{\ttilde_{i,j}^{(u)}\}_{u=1}^{\ktilde_{i,j}}$ is then defined
as
\begin{equation}
    \{\ttilde_{i,j}^{(u)}\}_{u=1}^{\ktilde_{i,j}}
    = \bigcup_{(x,y)\in\Neighborhood_{i,j}} \{t_{x,y}^{(\ell)}\}_{\ell=1}^{k_{x,y}},
\end{equation}
where $\ktilde_{i,j}$ is the new detection count for the superpixel at $(i,j)$.
In this way, the algorithm searches a small local area and adaptively borrows from pixels that are similar in both transverse position and reflectivity.

Once superpixel vectors have been formed, the windowing process of Section~\ref{sec:windowing} and the reflectivity estimation of Section~\ref{sec:refl_est} are repeated. 
The windowing is performed on the set of superpixel
detections $\{\ttilde_{i,j}^{(u)}\}_{u=1}^{\ktilde_{i,j}}$, resulting in a different (usually larger)
value of $\Nmax$.
Note that the $\Ncl$ computation and the reflectivity estimate change to account for the number of pixels $\Nsp$ contributing to the superpixel vector.

Ideally, the smallest possible $\Nsp$ such that $\Nmax\geq \Ncl$ would be chosen at each pixel, which would ensure accurate depth estimates with the minimum amount of spatial smoothing.
This could be accomplished by incorporating detections from one pixel at a time and re-windowing to check whether the $\Ncl$ criterion had been met. 
Unfortunately, this
repeated windowing of new detection vectors is too computationally intensive for large images.
Instead, we take a coarser approach that gradually increases the candidate neighborhood for forming superpixels by incrementing $\dsp$.
We cycle through the procedures of windowing, estimating reflectivity, and forming superpixels, gradually increasing $\dsp$ with each iteration from $\dsp = 0$ until either $\Nmax\geq \Ncl$ for all $(i,j)$ or some terminal neighborhood size $\dspmax$ has been reached. 
For any remaining pixels without a reliable depth estimate, $\hat{z}_{i,j}$ is filled in by inpainting during the depth estimation procedure.

\subsection{Depth Estimation}
It is assumed that all detections retained in $V_{i,j}$ are due to signal, although if too many noise clusters are falsely accepted, further rank-ordered mean (ROM) censoring as in \cite{Shin2015} can be useful in cleaning up the data.
The negative log-likelihood of the depth $z_{i,j}$ given only signal detections is
\begin{equation}
    \mathcal{L}_z\!\left(z_{i,j};\{\ttilde_{i,j}^{(\ell)}\}_{\ell \in V_{i,j}}\right)
    = - \sum_{\ell \in V_{i,j}} \log[s(\ttilde_{i,j}^{(\ell)}- 2z_{i,j}/c)].
\end{equation}
Again applying a smoothness penalization appropriate for depth maps of natural scenes, the PML depth estimate is 
\begin{align}
    \hat{\bz}^{\rm PML} = &\argmin_{\bz \, : \, z_{i,j}\in [0,\zmax)} \sum_{i=1}^{N_i} \sum_{j=1}^{N_j} \mathcal{L}_z\!\left(z_{i,j};\{t_{i,j}^{(\ell)}\}_{\ell \in V_{i,j}}\right) \nonumber \\
    &+ \beta_z \pen_z(\bz),
\end{align}
where $\beta_z \in [0,\infty)$ controls the amount of penalization.

\section{Results}\label{sec:results}
A detailed account of the experimental setup and procedure is given in~\cite{FPI2014} and its supplement~\cite{FPI2014supp}.
The important quantities for our algorithm are the illumination pulse width, measured to be $\Tp = 270$ ps, and the pulse repetition period $\Tr = 100$ ns. 
The SPAD detector quantum efficiency was $\eta = 0.35$.

In~\cite{FPI2014} and~\cite{Shin2015}, the photon-efficient methods are compared to ``ground truth'' reconstructions of reflectivity and depth, generated using conventional LIDAR processing on data from long acquisition times.
While these measurements serve as effective baseline comparisons, they still suffer from the same shortcomings as all LIDAR data. 
In particular, the conventional processing assumes only one depth exists at each point in the image, and we make this assumption as well. 
Taking into account multiple depths at a single pixel as in \cite{Shin2016multidepth} would require adjustments to our algorithm, since superpixels would borrow detections from multiple true depths, only one of which would be registered.
Since experimental LIDAR data has effects of shadowing or reflections from multiple depths, we consider the conventional processing to produce ``baseline'' estimates, but not ground truth.

\subsection{Simulated Results}
\newlength{\artheight}  
\settoheight\artheight{\begin{minipage}{ 0.14\linewidth}       \includegraphics[width=\textwidth]{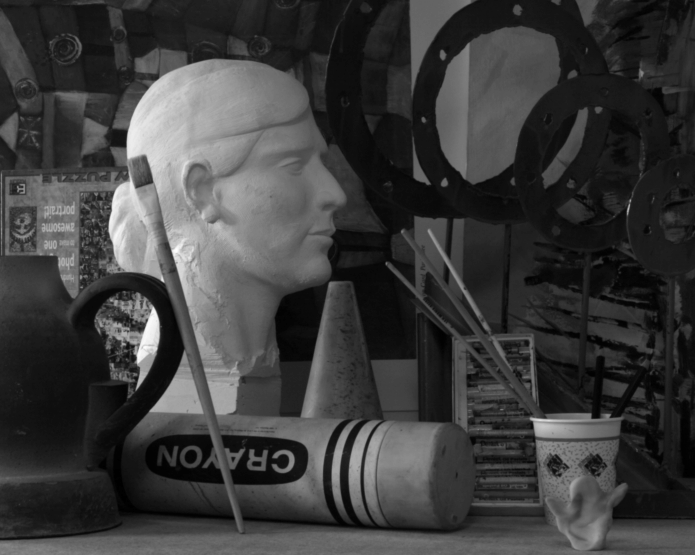}\end{minipage}}

\newlength{\bowlheight}  
\settoheight\bowlheight{\begin{minipage}{ 0.14\linewidth}       \includegraphics[width=\textwidth]{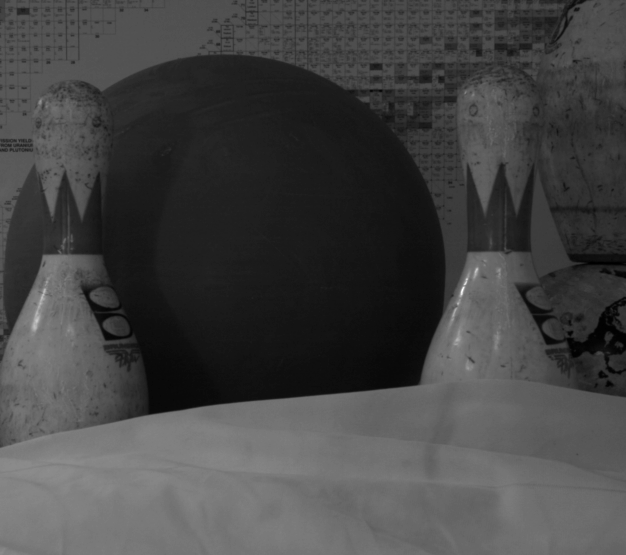}\end{minipage}}

\begin{figure*}
\captionsetup[subfigure]{labelformat=empty}
\centering
    \begin{minipage}{0.03\textwidth}
        \begin{sideways}
        \text{Reflectivity}
        \end{sideways}
    \end{minipage}
    \begin{minipage}{ 0.14\linewidth}
        \centering
        \subcaption{Ground truth}
        \includegraphics[width=\textwidth]{Art_m_refl_truth2}
        \small{\text{}}
    \end{minipage}
    \begin{minipage}{ 0.14\linewidth}
        \centering
        \subcaption{Photon count}
        \includegraphics[width=\textwidth]{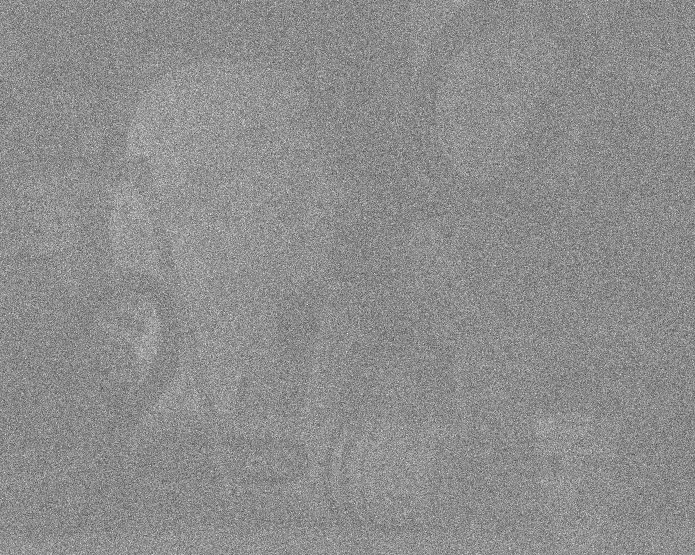}
        \small{\text{ }}
    \end{minipage}
    \begin{minipage}{ 0.14\linewidth}
        \centering
        \subcaption{Signal oracle}
        \includegraphics[width=\textwidth]{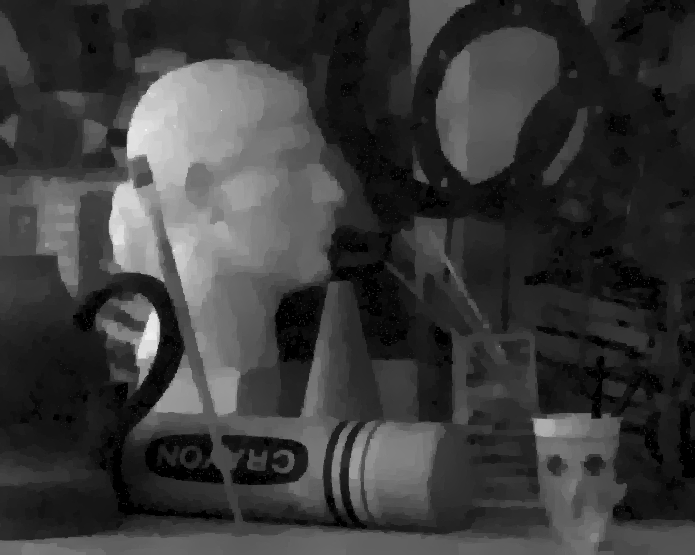}
        \small{\text{MSE = -69.5 dB}}
    \end{minipage}
    \begin{minipage}{ 0.14\linewidth}
        \centering
        \subcaption{\Shin}
        \includegraphics[width=\textwidth]{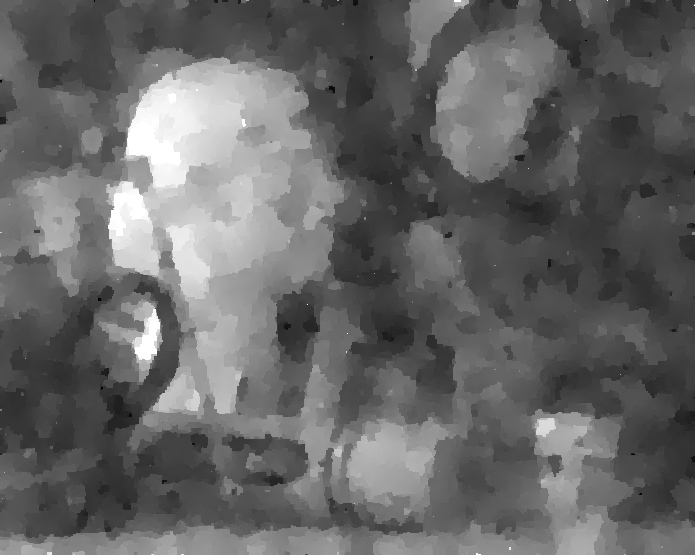}
        \small{\text{MSE = -56.4 dB}}
    \end{minipage}
    \begin{minipage}{ 0.14\linewidth}
        \centering
        \subcaption{Our method}
        \includegraphics[width=\textwidth]{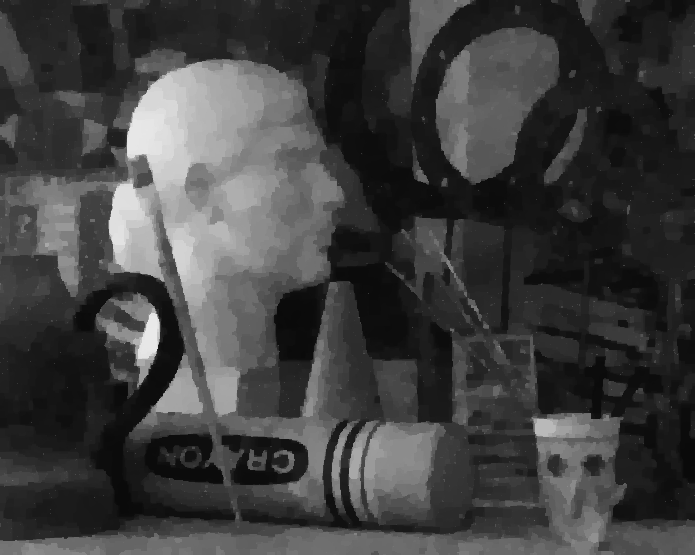}
        \small{\text{MSE = -69.0 dB}}
    \end{minipage}
    \begin{minipage}{0.03\linewidth}
        \subcaption{ }
        \includegraphics[ height=3.45\artheight]{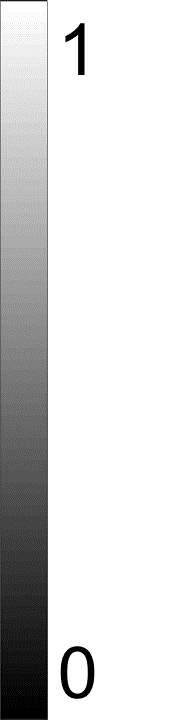}
        \small{\text{ }}
    \end{minipage}
    \begin{minipage}{ 0.14\linewidth}
        \centering
        \subcaption{Abs.\ error}
        \includegraphics[width=\textwidth]{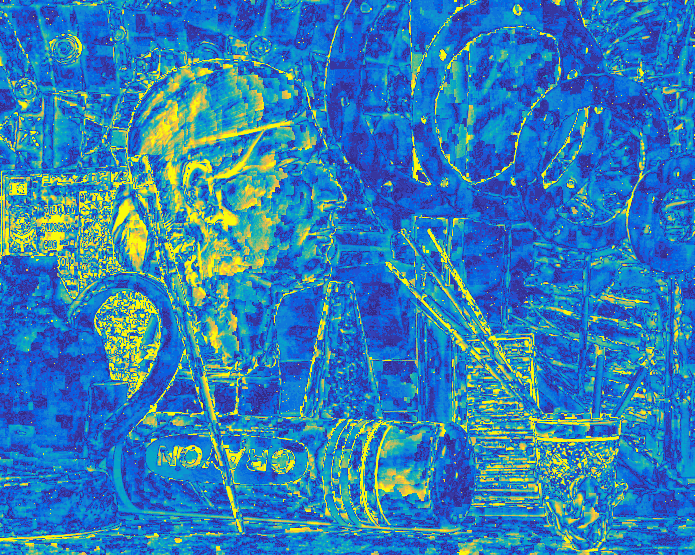}
        \small{\text{ }}
    \end{minipage}
    \begin{minipage}{0.03\linewidth}
        \subcaption{ }
        \includegraphics[ height=3.45\artheight]{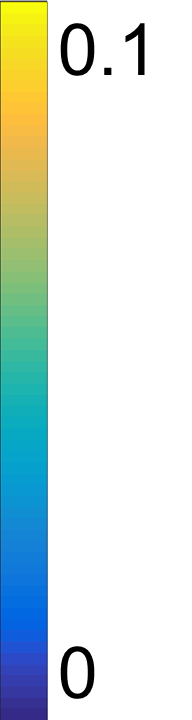}
        \small{\text{ }}
    \end{minipage}
    \vspace{1mm}

   \begin{minipage}{0.03\textwidth}
        \begin{sideways}
        \text{Reflectivity}
        \end{sideways}
    \end{minipage}
    \begin{minipage}{ 0.14\linewidth}
        \centering
        \includegraphics[width=\textwidth]{Bowling_m_refl_truth2}
        \small{\text{ }}
    \end{minipage}
    \begin{minipage}{ 0.14\linewidth}
        \centering
        \includegraphics[width=\textwidth]{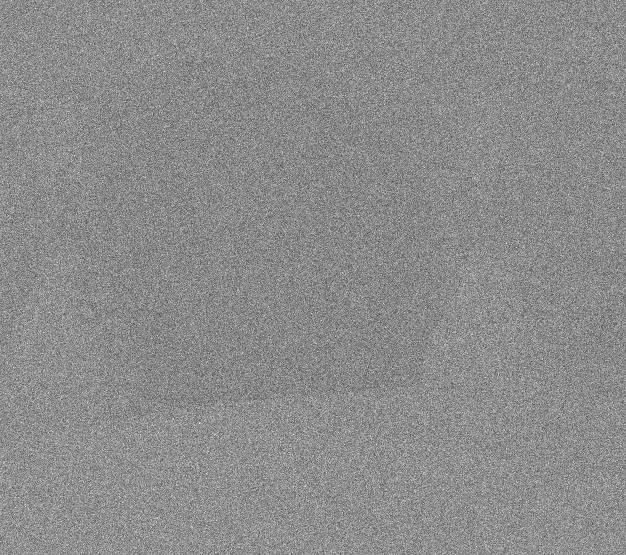}
        \small{\text{ }}
    \end{minipage}
    \begin{minipage}{ 0.14\linewidth}
        \centering
        \includegraphics[width=\textwidth]{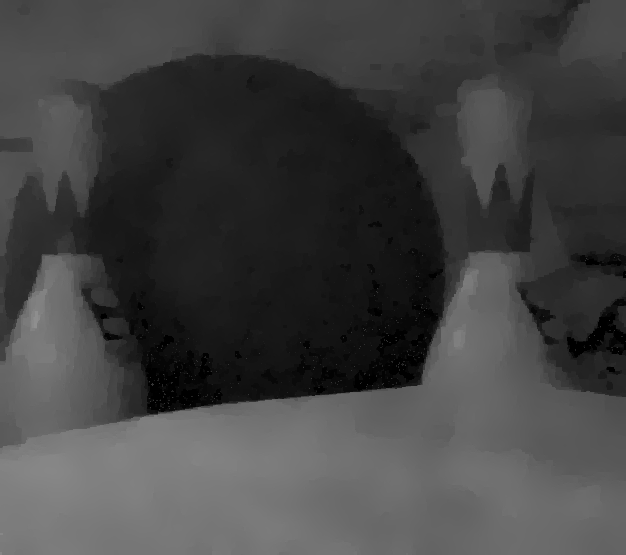}
        \small{\text{MSE = -74.8 dB}}
    \end{minipage}
    \begin{minipage}{ 0.14\linewidth}
        \centering
        \includegraphics[width=\textwidth]{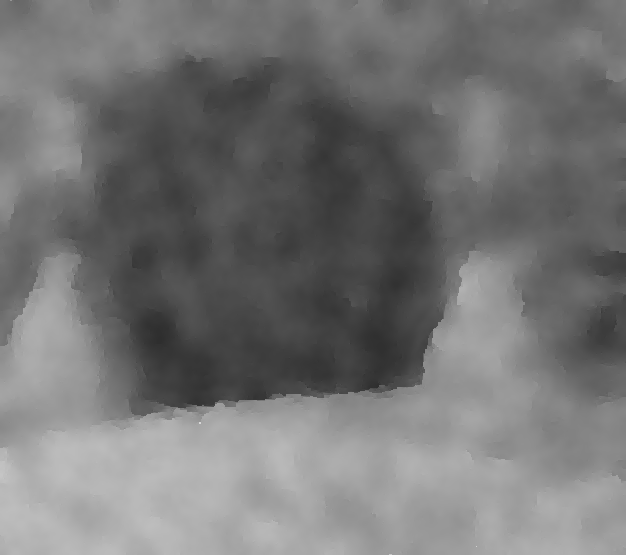}
        \small{\text{MSE = -56.9 dB}}
    \end{minipage}
    \begin{minipage}{ 0.14\linewidth}
        \centering
        \includegraphics[width=\textwidth]{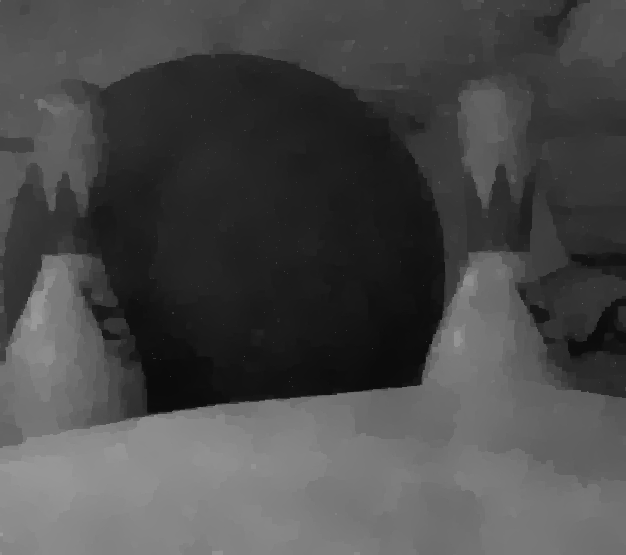}
        \small{\text{MSE = -71.6 dB}}
    \end{minipage}
    \begin{minipage}{0.03\linewidth}
        \includegraphics[ height=3.45\bowlheight]{Slide1}
        \small{\text{ }}
    \end{minipage}
    \begin{minipage}{ 0.14\linewidth}
        \centering
        \includegraphics[width=\textwidth]{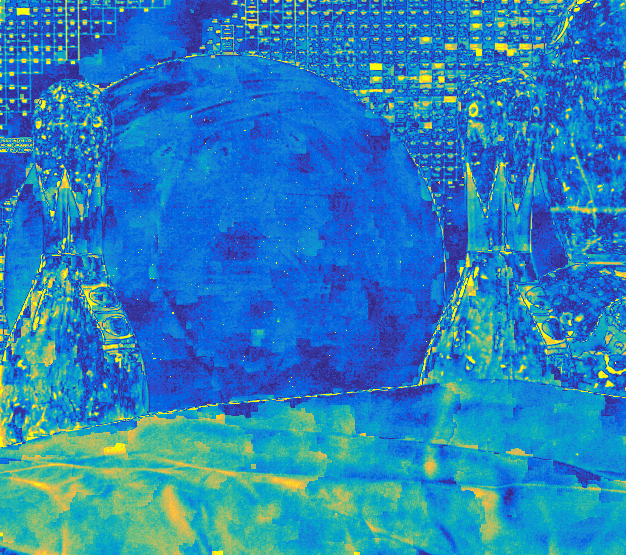}
        \small{\text{ }}
    \end{minipage}
    \begin{minipage}{0.03\linewidth}
        \includegraphics[ height=3.45\bowlheight]{Slide2}
        \small{\text{ }}
    \end{minipage}
    
    \vspace{5mm}

    \begin{minipage}{0.03\textwidth}
        \begin{sideways}
        \text{Depth}
        \end{sideways}
    \end{minipage}
    \begin{minipage}{ 0.14\linewidth}
        \centering
        \subcaption{Ground truth}
        \includegraphics[width=\textwidth]{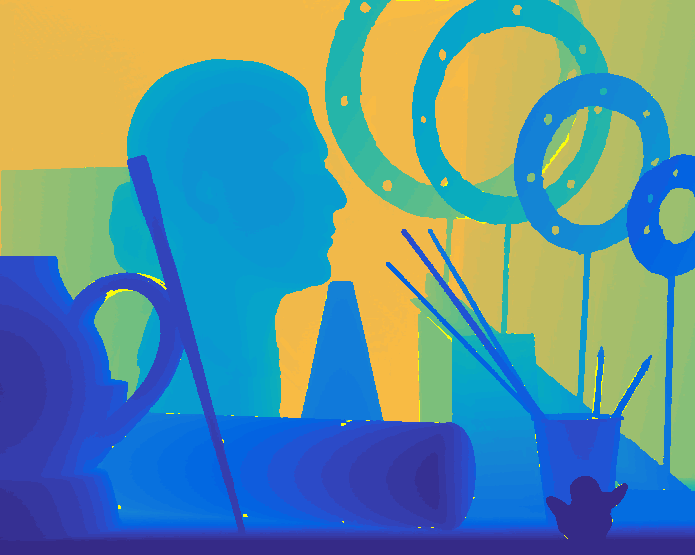}
        \small{\text{ }}
    \end{minipage}
    \begin{minipage}{ 0.14\linewidth}
        \centering
        \subcaption{Log-matched filter}
        \includegraphics[width=\textwidth]{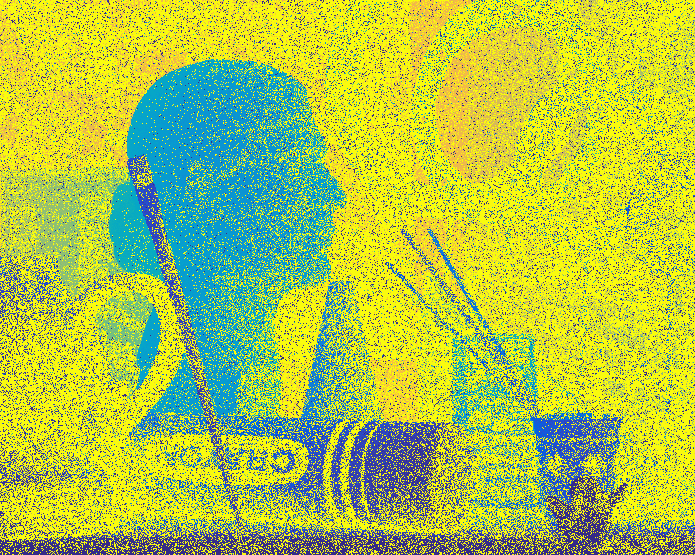}
        \small{\text{ }}
    \end{minipage}
    \begin{minipage}{ 0.14\linewidth}
        \centering
        \subcaption{Signal oracle}
        \includegraphics[width=\textwidth]{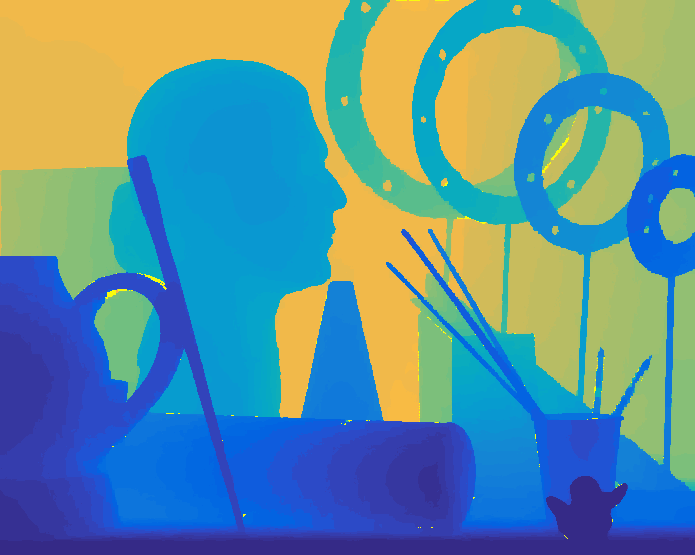} \\
        \small{\text{RMSE = 0.036 m}}
    \end{minipage}
    \begin{minipage}{ 0.14\linewidth}
        \centering
        \subcaption{\Shin}
        \includegraphics[width=\textwidth]{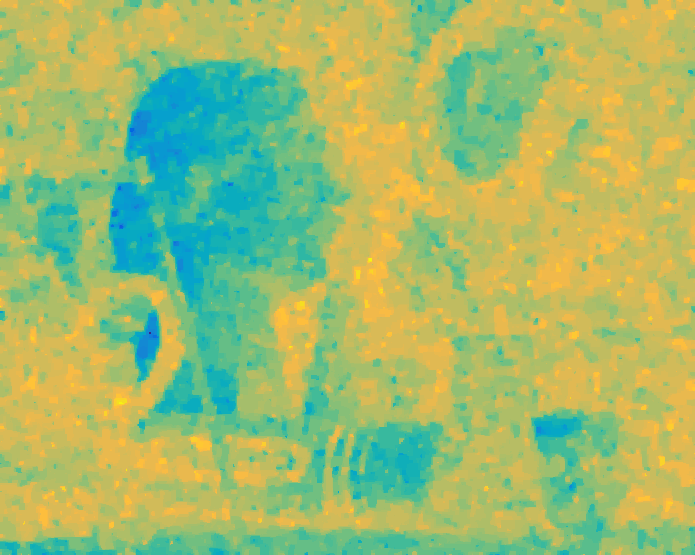}
        \small{\text{RMSE = 5.38 m}}
    \end{minipage}
    \begin{minipage}{ 0.14\linewidth}
        \centering
        \subcaption{Our method}
        \includegraphics[width=\textwidth]{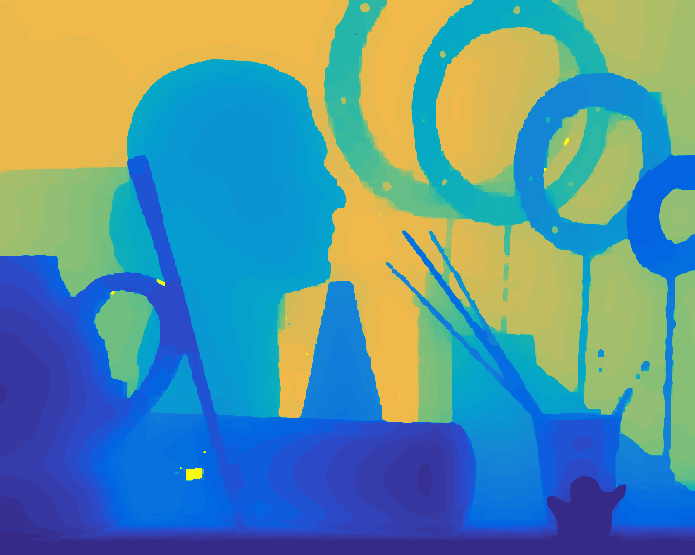}
        \small{\text{RMSE = 0.094 m}}
    \end{minipage}
    \begin{minipage}{0.03\linewidth}
        \subcaption{ }
        \includegraphics[ height=3.45\artheight]{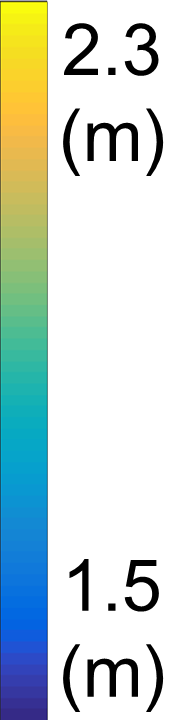}
        \small{\text{ }}
    \end{minipage}
    \begin{minipage}{ 0.14\linewidth}
        \centering
        \subcaption{Abs.\ error}
        \includegraphics[width=\textwidth]{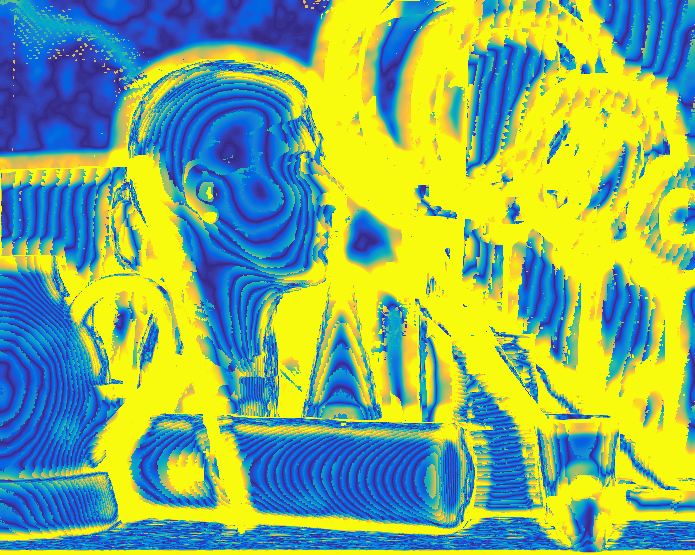}
        \small{\text{ }}
    \end{minipage}
    \begin{minipage}{0.03\linewidth}
        \subcaption{ }
        \includegraphics[ height=3.45\artheight]{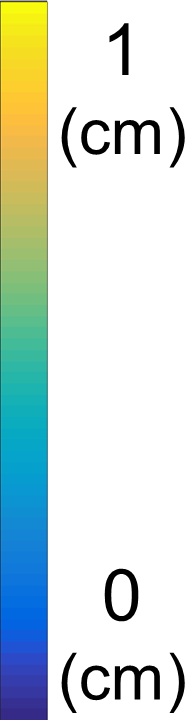}
        \small{\text{ }}
    \end{minipage}
   
    \vspace{1mm}

    \begin{minipage}{0.03\textwidth}
        \begin{sideways}
        \text{Depth}
        \end{sideways}
    \end{minipage}
    \begin{minipage}{ 0.14\linewidth}
        \centering
        \includegraphics[width=\textwidth]{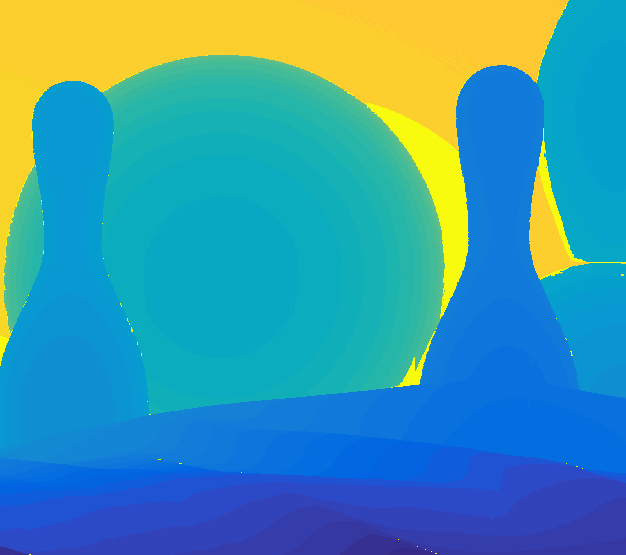}
        \small{\text{ }}
    \end{minipage}
    \begin{minipage}{ 0.14\linewidth}
        \centering
        \includegraphics[width=\textwidth]{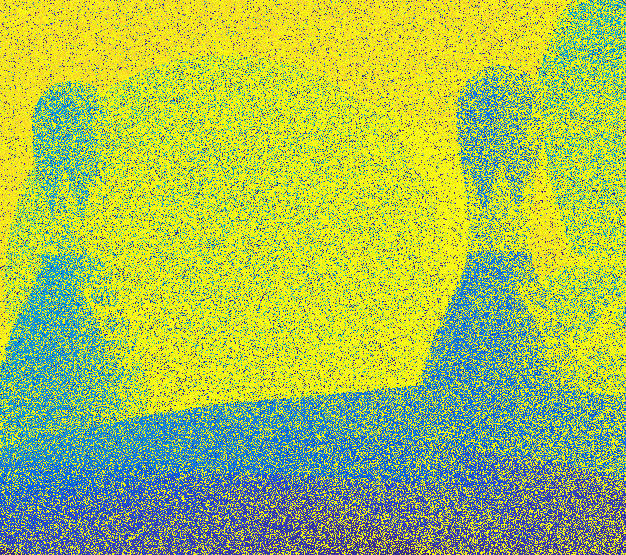}
        \small{\text{ }}
    \end{minipage}
    \begin{minipage}{ 0.14\linewidth}
        \centering
        \includegraphics[width=\textwidth]{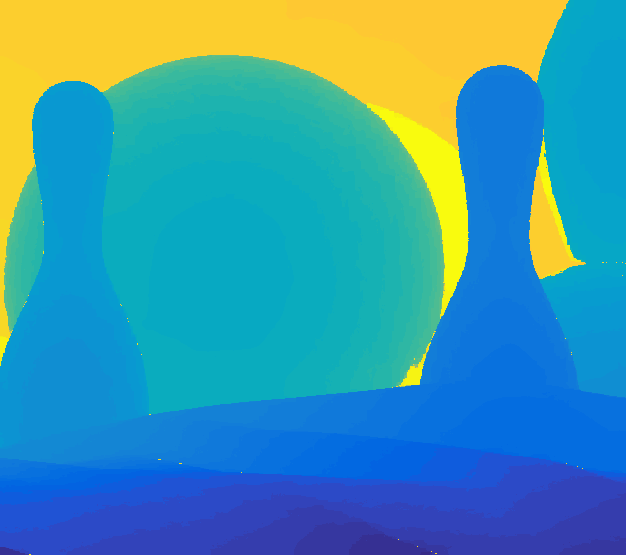} \\
        \small{\text{RMSE = 0.015 m}}
    \end{minipage}
    \begin{minipage}{ 0.14\linewidth}
        \centering
        \includegraphics[width=\textwidth]{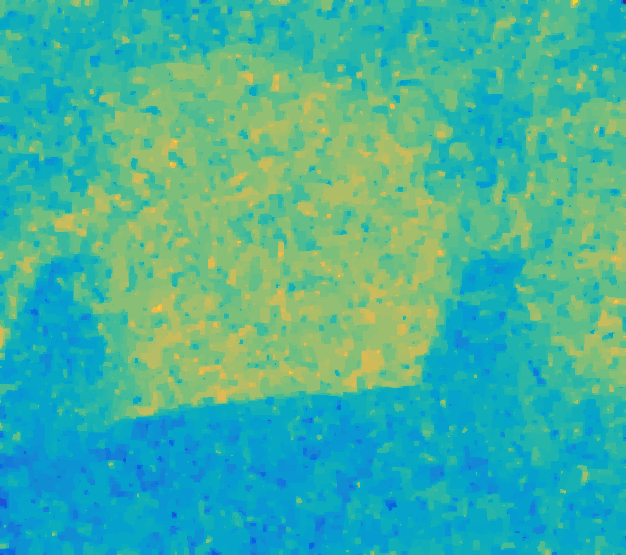}
        \small{\text{RMSE = 5.69 m}}
    \end{minipage}
    \begin{minipage}{ 0.14\linewidth}
        \centering
        \includegraphics[width=\textwidth]{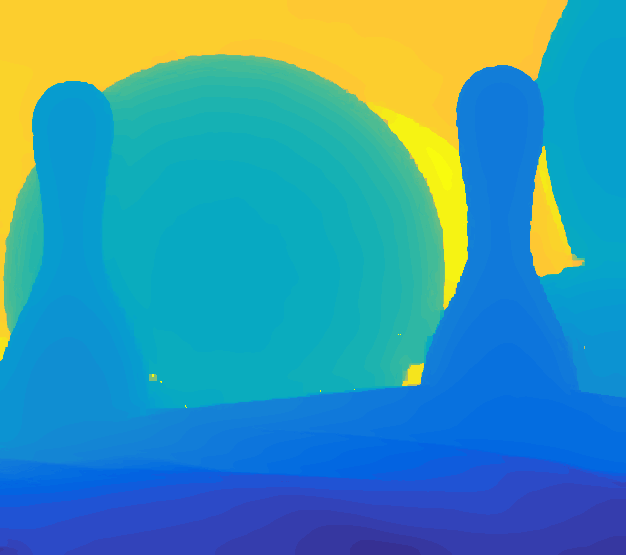}
        \small{\text{RMSE = 0.067 m}}
    \end{minipage}
    \begin{minipage}{0.03\linewidth}
        \includegraphics[ height=3.45\bowlheight]{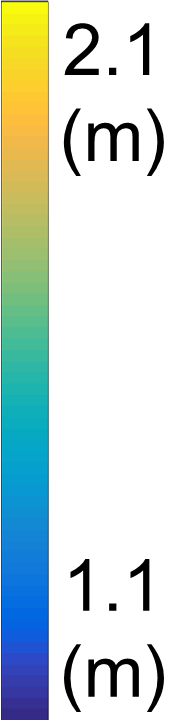}
        \small{\text{ }}
    \end{minipage}
    \begin{minipage}{ 0.14\linewidth}
        \centering
        \includegraphics[width=\textwidth]{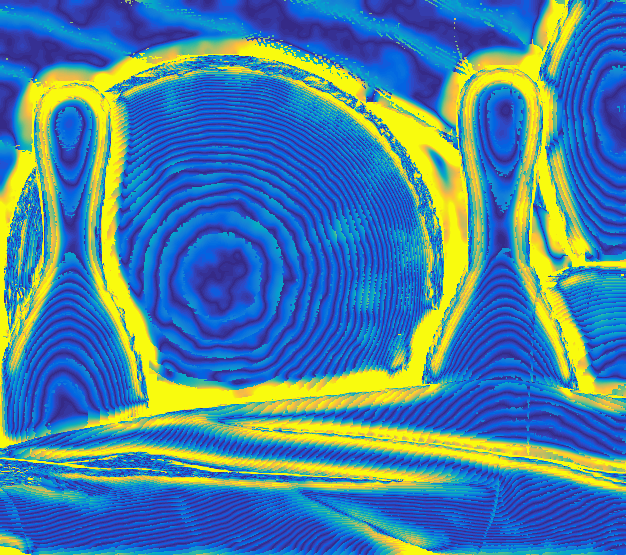}
        \small{\text{ }}
    \end{minipage}
    \begin{minipage}{0.03\linewidth}
        \includegraphics[ height=3.45\bowlheight]{Slide3}
        \small{\text{ }}
    \end{minipage}
    
    \caption{Simulated processing results for Art and Bowling scenes~\cite{scharstein2007} at SBR = 0.04 and 2.0 signal photons per pixel on average. 
    Note that the depth estimates with the method of \cite{Shin2015} are completely out of range of the actual scene and are instead shown for the range of 6 to 8 meters.}
    \label{fig:sim_results}
    
\end{figure*}

In order to quantify the algorithm performance compared to an actual ground-truth reference, we first simulated data sets using the model outlined in Section~\ref{sec:measurement}, where each pixel has only a single true depth.
The same parameters from the experiments were used in the simulation in order to maintain consistency.
Furthermore, although detections are generated from a model, we use real scenes from the Middlebury dataset \cite{scharstein2007} to form $\balpha$ and $\bz$.
In particular, we chose the Art and Bowling scenes as representative of fairly complex and fairly simple scenes, respectively.
The Art scene is $695 \times 555$ pixels, and the Bowling scene is $626 \times 555$ pixels.

Signal counts were generated as Poisson random variables with parameters equal to scaled pixel intensities.
Signal detection times were generated from a Gaussian pulse shape with mean $\bz_{i,j}$ and $\sigma = \Tp/2$.
Background detection counts were also generated as Poisson random variables, and given the count at each pixel, the detection times were generated as uniform random variables over the repetition period $[0,\Tr)$.
In order to meet the low-flux requirement, scenes were simulated so that the average pixel would require 500 illuminations to generate one signal photon. 
Thus, performance evaluation of scenes with 2.0 and 3.0 signal ppp used 1000 and 1500 illumination periods per pixel, respectively. 
At the maximum evaluated noise level (with SBR = 0.04), the average photon detection rate was one detection in approximately 5$\%$ of illumination periods.

To quantify performance, we use the mean-squared error (MSE) in dB for reflectivity:
\begin{equation*}
\MSE (\balpha,\hat{\balpha}) = 10 \log_{10}\left (\frac{1}{n^2} \sum_{i=1}^{N_i} \sum_{j=1}^{N_j} (\alpha_{i,j} - \hat{\alpha}_{i,j} )^2\right ),
\end{equation*}
and the root mean-square error for depth:
\begin{equation*}
\RMSE (\bz,\hat{\bz}) = \sqrt{\frac{1}{n^2} \sum_{i=1}^{N_i} \sum_{j=1}^{N_j} (z_{i,j} - \hat{z}_{i,j} )^2}.
\end{equation*}

Figure~\ref{fig:sim_results} shows example simulation results for both scenes at SBR = 0.04 and only 2.0 signal photons per pixel on average.
The scaled photon count is shown for reflectivity and the log-matched filter output for depth as a baseline for what conventional methods produce for such noisy, photon-efficient data.
We compare the results of our proposed method with that of {\ShinEtAl}~\cite{Shin2015}, which is the state-of-the-art for photon-efficient imaging at lower noise levels.
We also show the ideal results from a \emph{signal oracle}, which represents the ideal case of perfect unmixing and using only the signal detections for estimation (equivalently, SBR = $\infty$).

Throughout the simulations, we use $\dspmax = 3$, $\tausp$ = 0.05, and $\tauFA = 0.01$ for our algorithm, which work for a variety of scenes and experimental conditions. 
These parameters were mainly tuned for very-low SBR data (around 25 times as much background as signal) and could be adjusted to optimize performance for different noise conditions or a particular scene.

The results in Figure~\ref{fig:sim_results} exemplify the typical performance of the different methods.
For reflectivity, it is clear that high levels of background reduce contrast too much for the method of \cite{Shin2015} to produce a good estimate from detection counts alone.
The unmixing does a much better job at estimating the number of signal detections at each pixel.
In particular, the absolute error maps show the smallest errors for the darkest regions, since formation of superpixels allows for precise fractional estimates of signal photon counts in these areas.

In the case of depth estimation, the method of \cite{Shin2015} fails, as noise detections pull the depth estimates towards the mean scene depth (7.5 meters in this case).
For our unmixing method, the windowing procedure is much more effective at handling the high-variance noise. The largest errors that remain in our depth estimation occur in the darkest regions of the scene, particularly at object boundaries.
In many of the dark regions, forming superpixels is enough to overcome the low signal photon count.
At object boundaries, however, $\Nsp$ decreases since many candidate pixels in the neighborhood fall outside the reflectivity tolerance so the signal clusters are too small, or if the reflectivity contrast between objects at different depths is not sufficient, the superpixels will borrow pixels at multiple depths, causing errors.
Nevertheless, the unmixing process produces depth estimates that are almost as good as the signal oracle in many cases.

\begin{figure}
\centering
    \begin{subfigure}{\linewidth}
        \centering
        \includegraphics[trim={0 14.5cm 0 0},clip,width=0.9\textwidth]{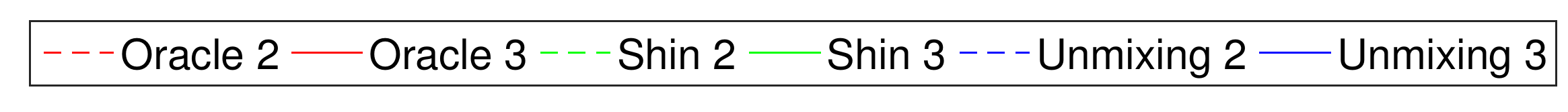}
    \end{subfigure}
    \\
    \begin{subfigure}{0.49\linewidth}
        \centering
        \includegraphics[width=\textwidth]{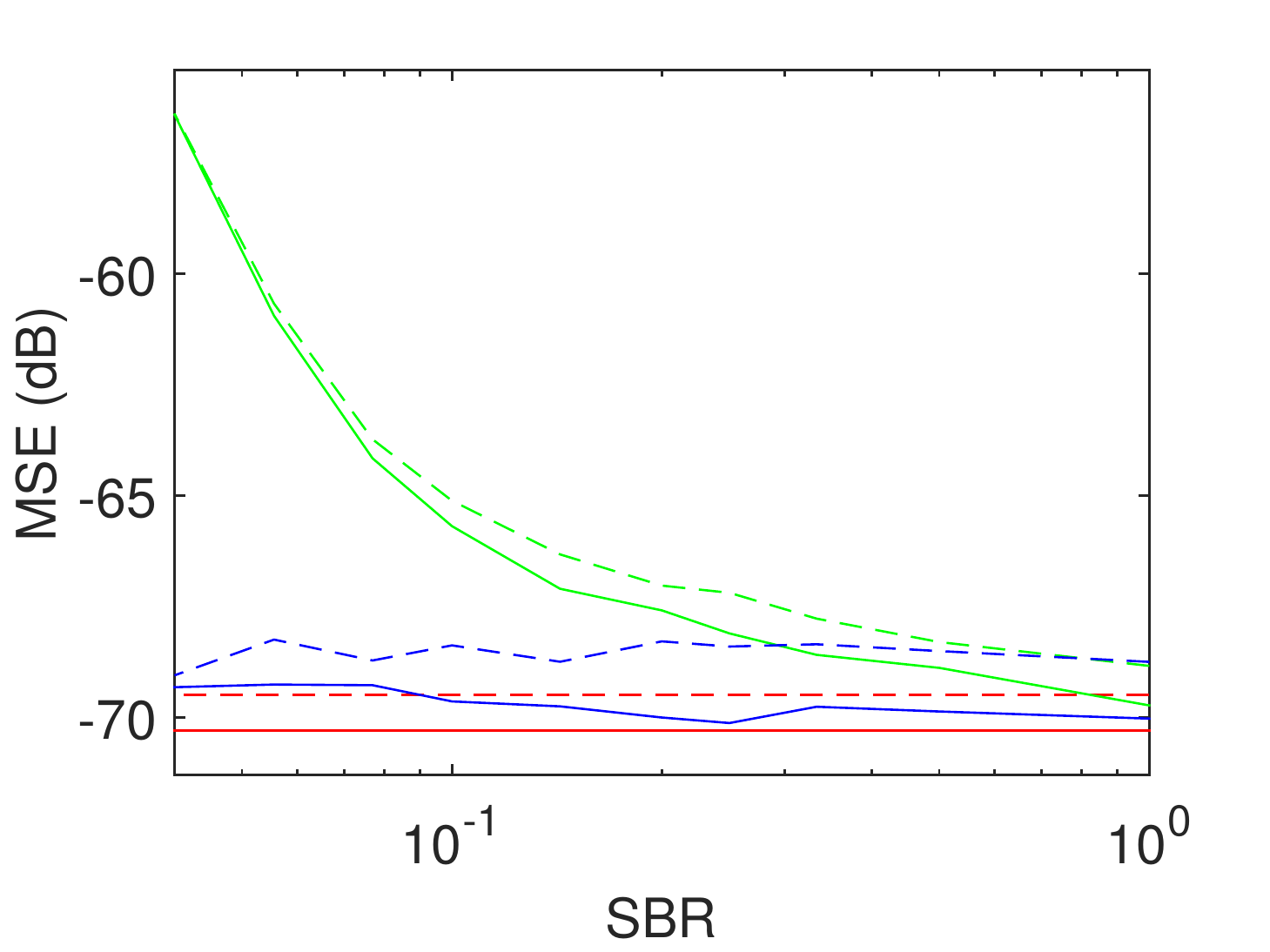}
        \caption{Art reflectivity}
    \end{subfigure}
    \begin{subfigure}{0.49\linewidth}
        \centering
        \includegraphics[width=\textwidth]{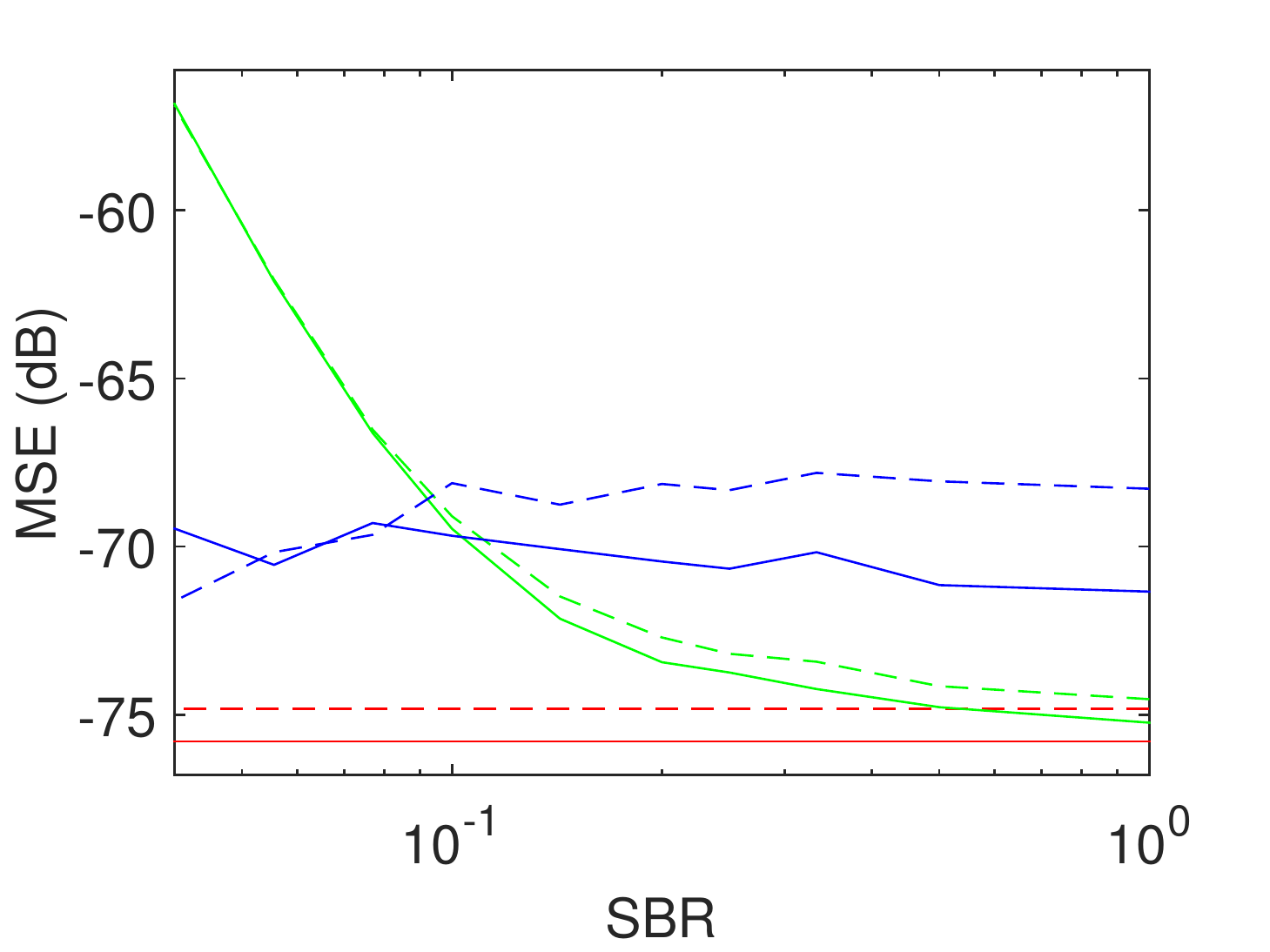}
        \caption{Bowling reflectivity}
    \end{subfigure}
    \\
    \begin{subfigure}{0.49\linewidth}
        \centering
        \includegraphics[width=\textwidth]{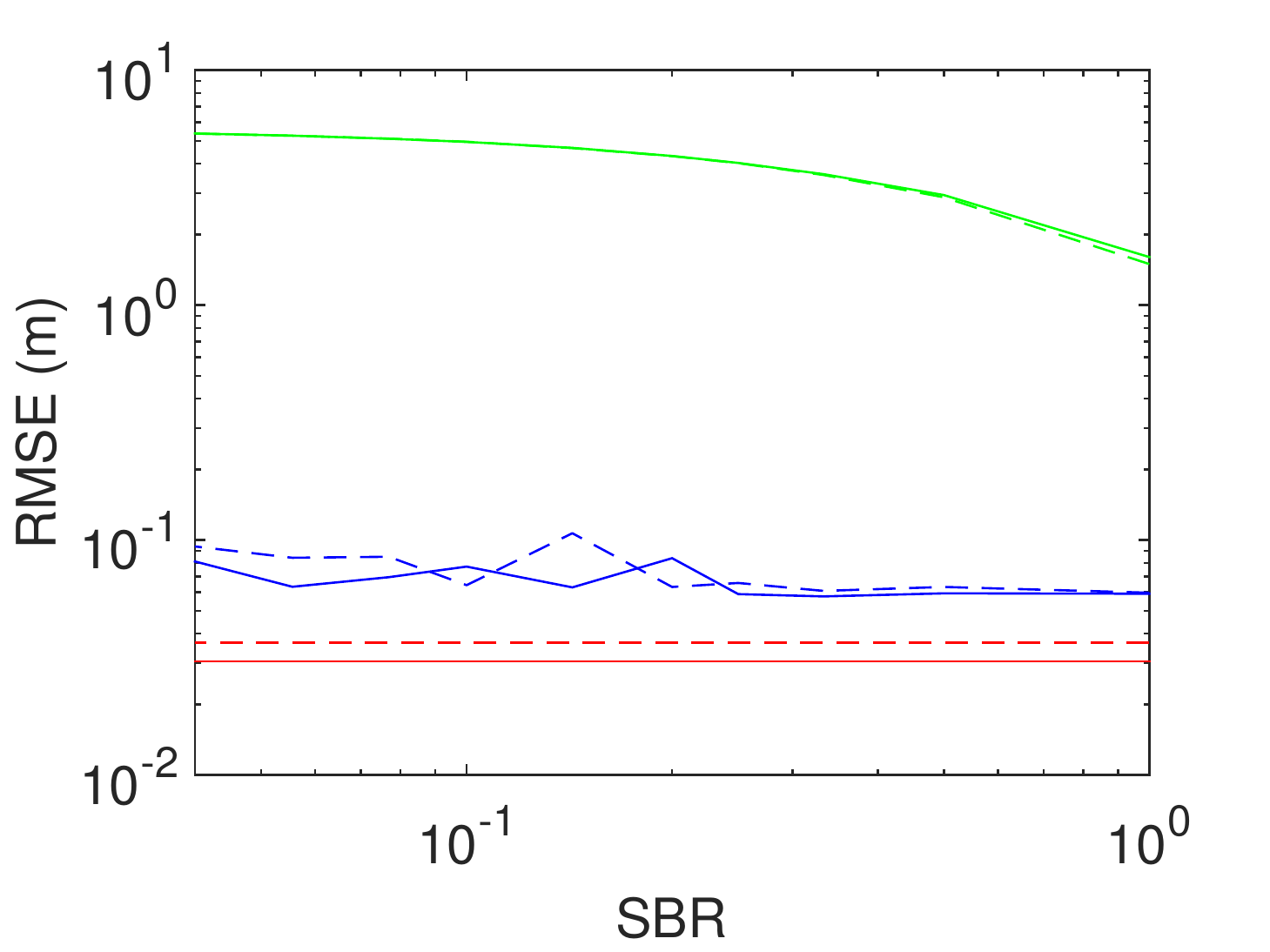}
        \caption{Art depth}
    \end{subfigure}
    \begin{subfigure}{0.49\linewidth}
        \centering
        \includegraphics[width=\textwidth]{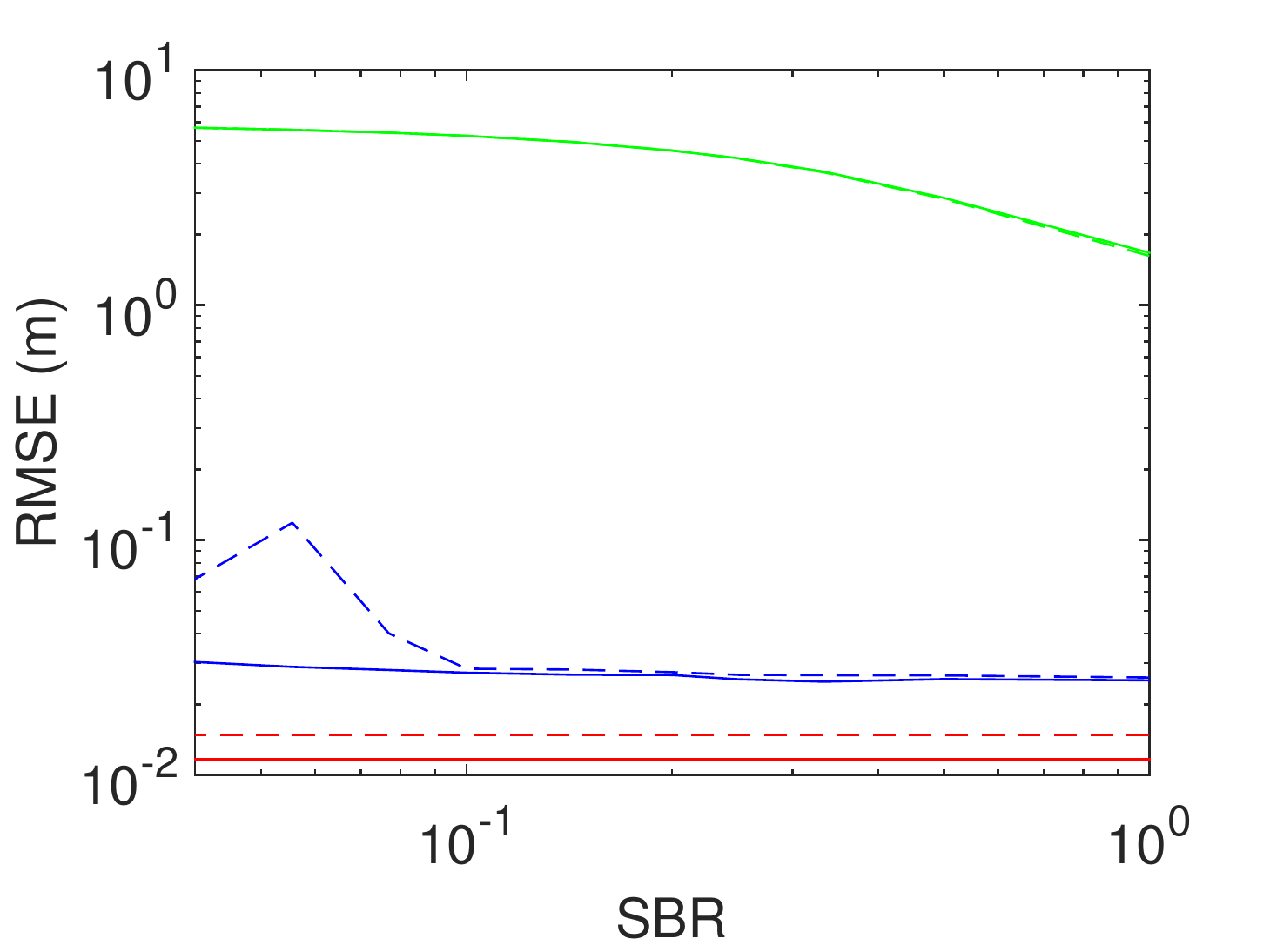}
        \caption{Bowling depth}
    \end{subfigure}
    \caption{Performance evaluation for reflectivity and depth estimation for simulated data sets with 2.0 and 3.0 signal photon detections per pixel on average and a range of SBR values. \emph{Oracle} refers to performance of a penalized ML estimator using only signal detections (SBR = $\infty$). Plotted performance is the average error of 10 trials for each value of SBR\@.
    }
    \label{fig:performance_results}
\end{figure}

Figure~\ref{fig:performance_results} contains plots comparing the oracle, {\Shin}, and unmixing methods for 2.0 and 3.0 signal detections per pixel at various SBR levels. 
The MSE and RMSE metrics are shown for the $\beta_\alpha$ and $\beta_z$ values that produced the best average performance over 10 trials at each value of SBR\@.
As expected, the best performance is achieved by the oracle estimator with the most signal detections per pixel, since this case has the most signal information available and is not corrupted by background.
Estimation of both parameters improves in general for all methods as the signal detection count increases.
It is also clear that the reflectivity and depth estimation performance of {\Shin} degrades significantly as SBR decreases.
This is due to the shortcomings of the binomial estimator for reflectivity and the limitations of the ROM censoring for removing noise detections at low SBR\@.
For our proposed unmixing method, the parameter estimation performance also tends to increase as SBR decreases, although the change in error is smaller than for {\Shin}, indicating a higher robustness to noise.
At SBR = 0.04, our reflectivity estimate outperforms the method of \cite{Shin2015} by about 15 dB\@.
The difference in depth estimation error is even more stark---at SBR = 0.04, our method has RMSE almost two orders of magnitude better than {\Shin}.

\subsection{Experimental Results}
\newlength{\manheight} 
\settoheight\manheight{\begin{minipage}{ 0.14\linewidth}       \includegraphics[width=\textwidth]{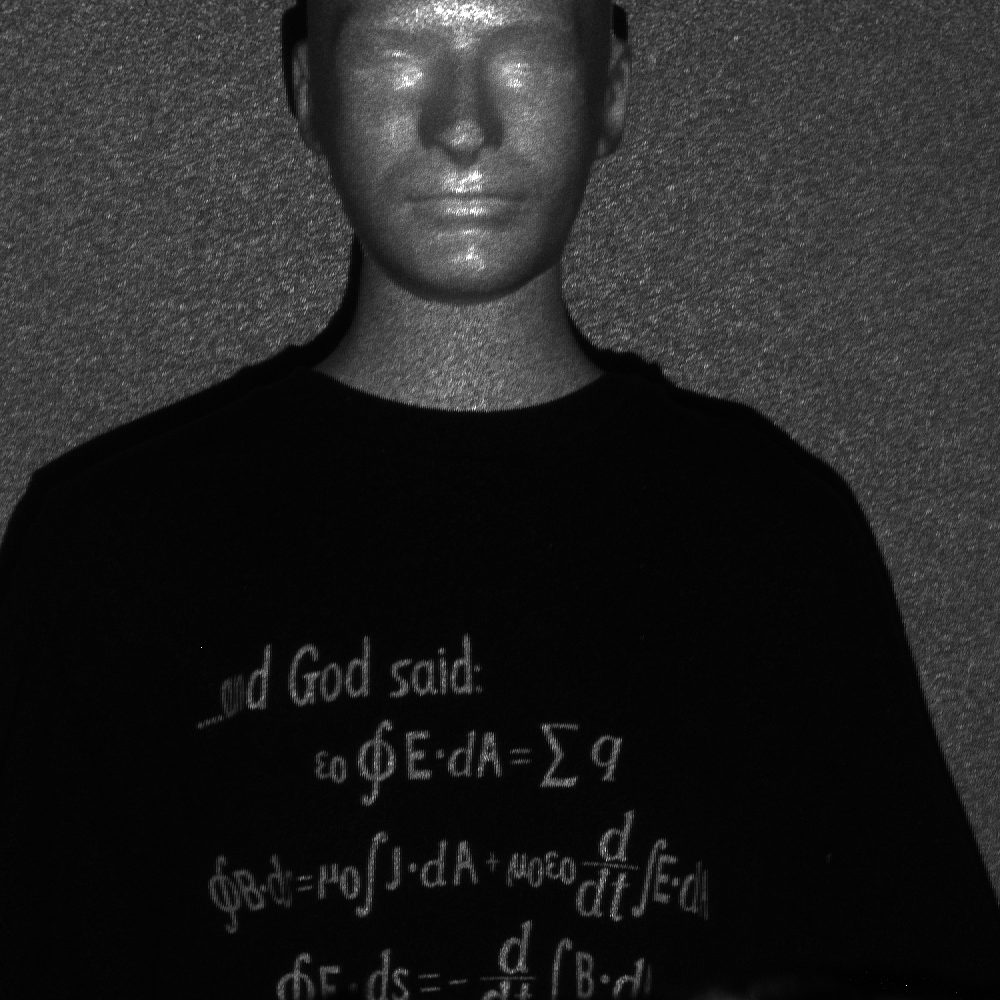}\end{minipage}}

\begin{figure*}
\captionsetup[subfigure]{labelformat=empty}
\centering
    \begin{minipage}{0.03\textwidth}
        \begin{sideways}
        \text{Reflectivity}
        \end{sideways}
    \end{minipage}
    \begin{minipage}{ 0.14\linewidth}   
        \centering
        \subcaption{Baseline}
        \includegraphics[width=\textwidth]{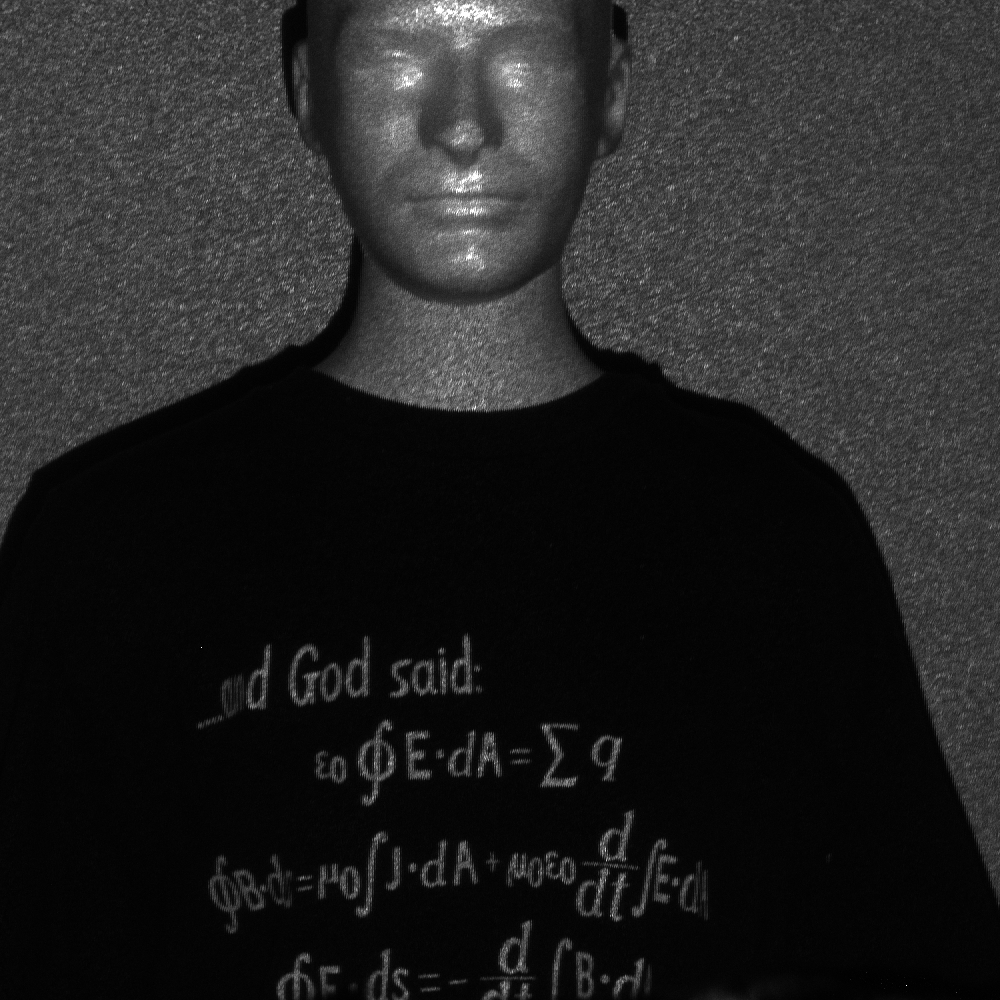}
        \small{\text{ }}
    \end{minipage}
    \begin{minipage}{ 0.14\linewidth}   
        \centering
        \subcaption{Photon count}
        \includegraphics[width=\textwidth]{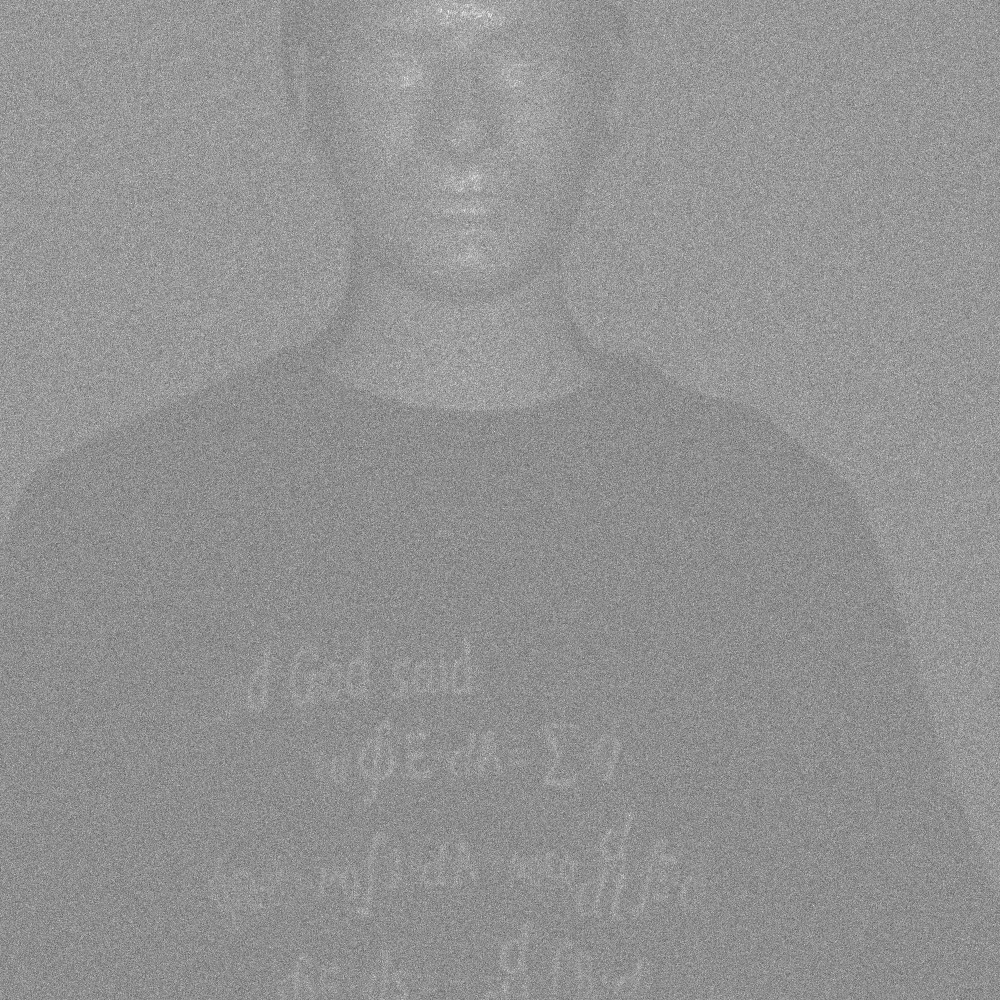}
        \small{\text{ }}
    \end{minipage}
    \begin{minipage}{ 0.14\linewidth}
        \centering
        \subcaption{Signal oracle}
        \includegraphics[width=\textwidth]{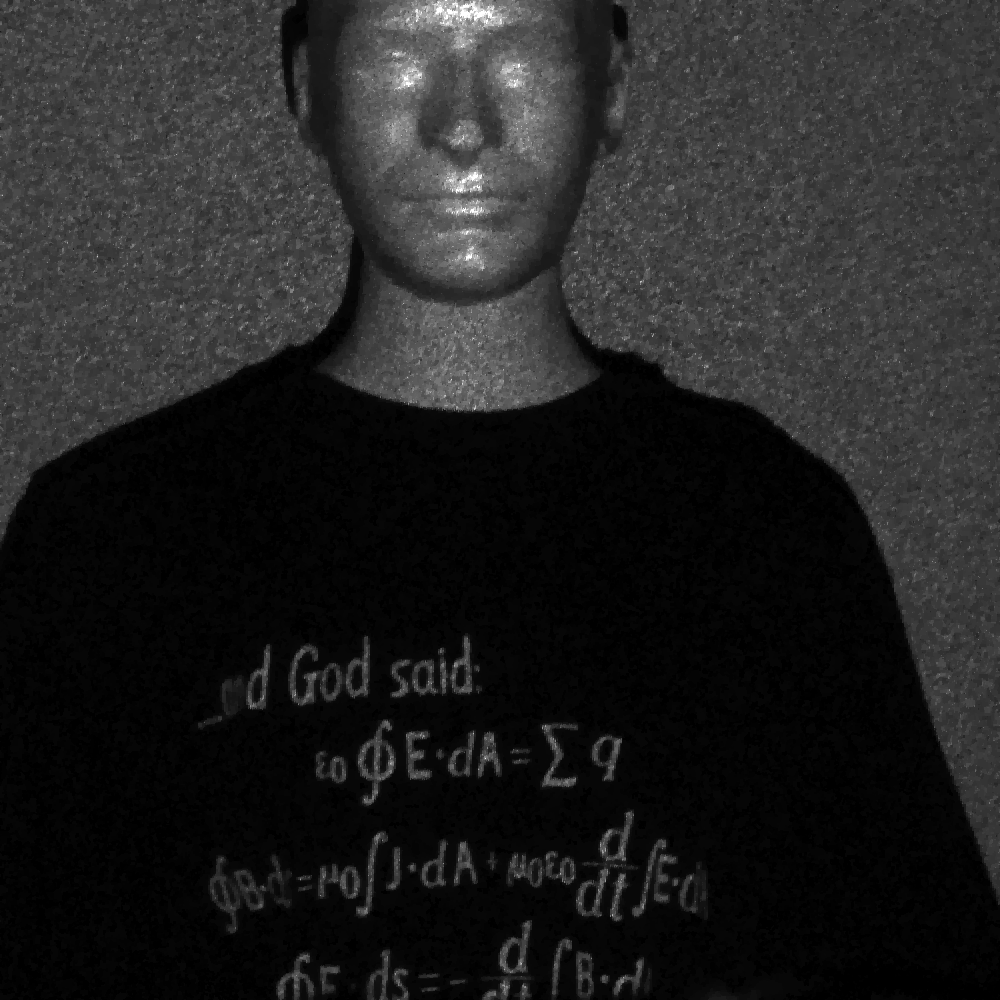}
        \small{\text{MSE = -58.7 dB}}
    \end{minipage} 
    \begin{minipage}{ 0.14\linewidth}
        \centering
        \subcaption{\Shin}
        \includegraphics[width=\textwidth]{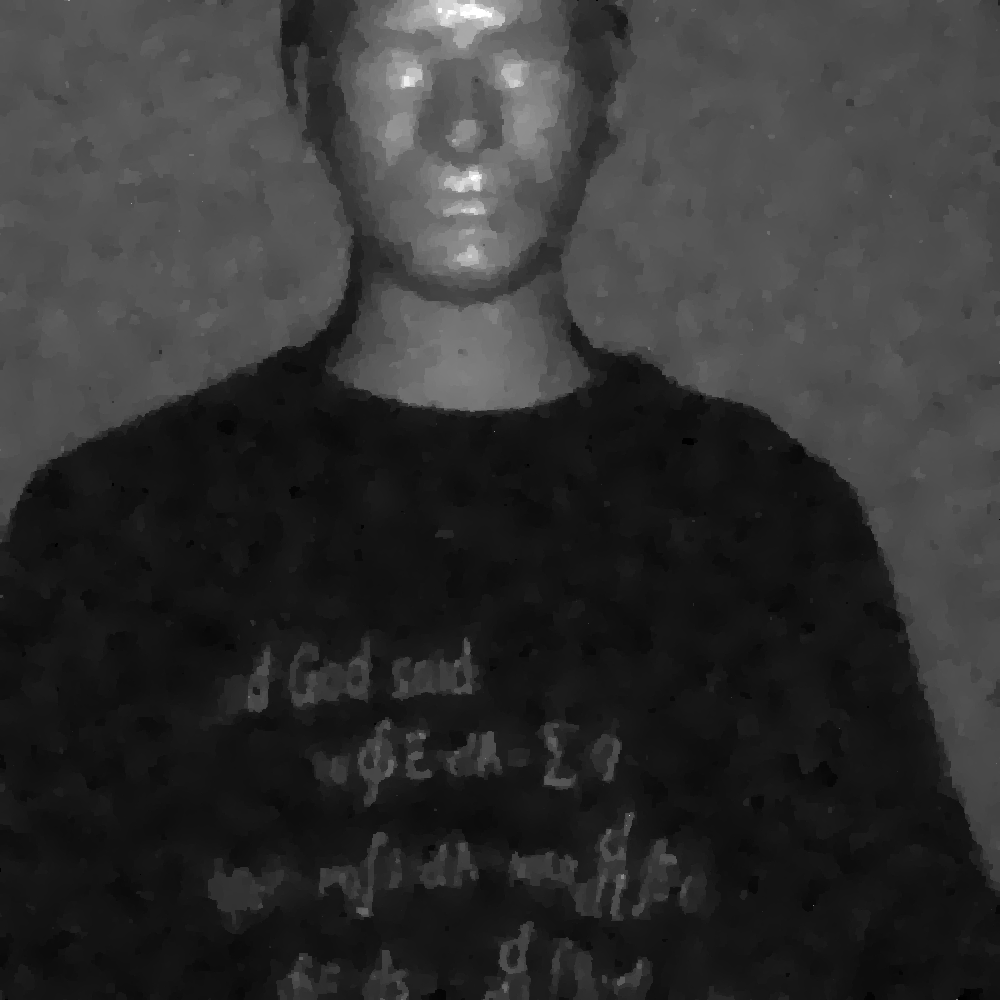}
        \small{\text{MSE = -57.4 dB}}
    \end{minipage} 
    \begin{minipage}{ 0.14\linewidth}
        \centering
        \subcaption{Our method}
        \includegraphics[width=\textwidth]{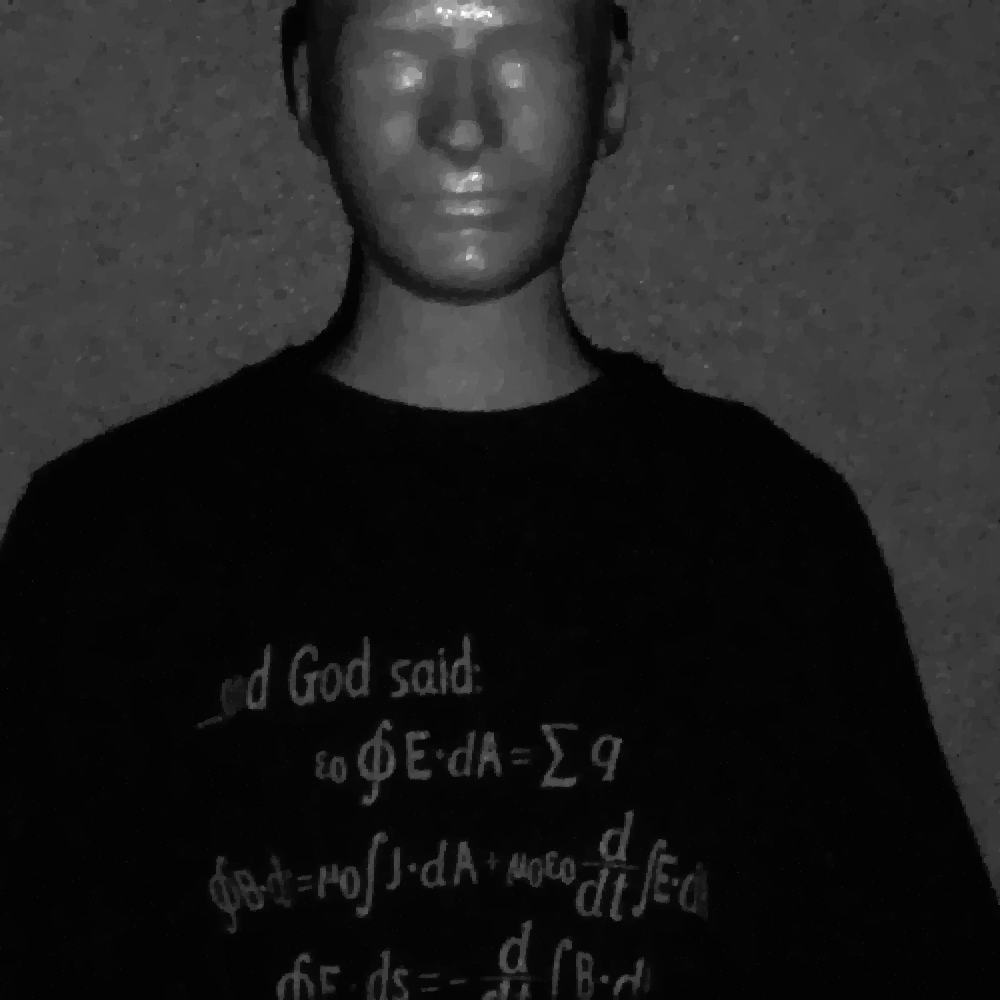}
        \small{\text{MSE  = -58.5 dB}}
    \end{minipage} 
    \begin{minipage}{0.03\linewidth}
        \centering
        \subcaption{ }
        \includegraphics[ height=3.5\manheight]{Slide1}
        \small{\text{ }}
    \end{minipage}
    \begin{minipage}{ 0.14\linewidth}   
        \centering
        \subcaption{Abs.\ error}
        \includegraphics[width=\textwidth]{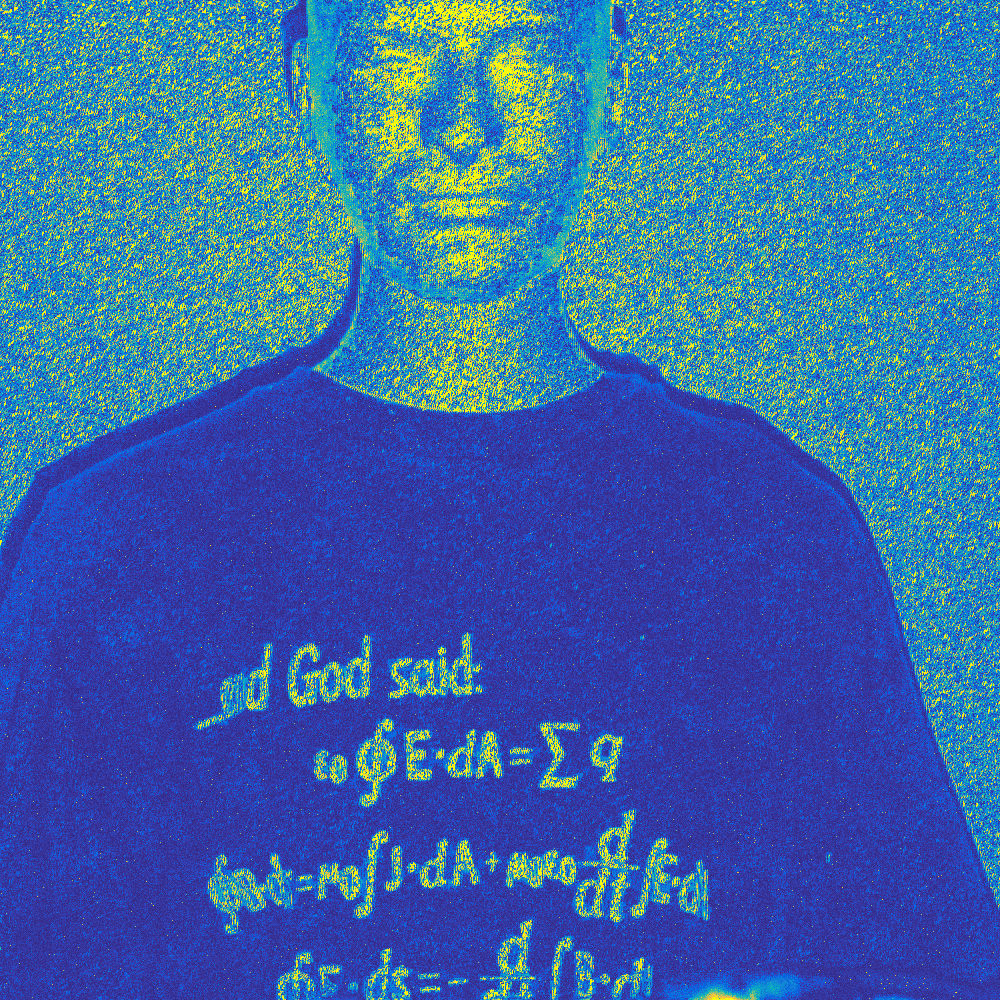}
        \small{\text{ }}
   \end{minipage}
   \begin{minipage}{0.03\linewidth}
        \centering
        \subcaption{ }
        \includegraphics[ height=3.5\manheight]{Slide2}
        \small{\text{ }}
    \end{minipage}
   
   \vspace{1mm}
   
    \begin{minipage}{0.03\textwidth}
        \begin{sideways}
        \text{Depth}
        \end{sideways}
    \end{minipage}
    \begin{minipage}{ 0.14\linewidth}   
        \centering
        \includegraphics[width=\textwidth]{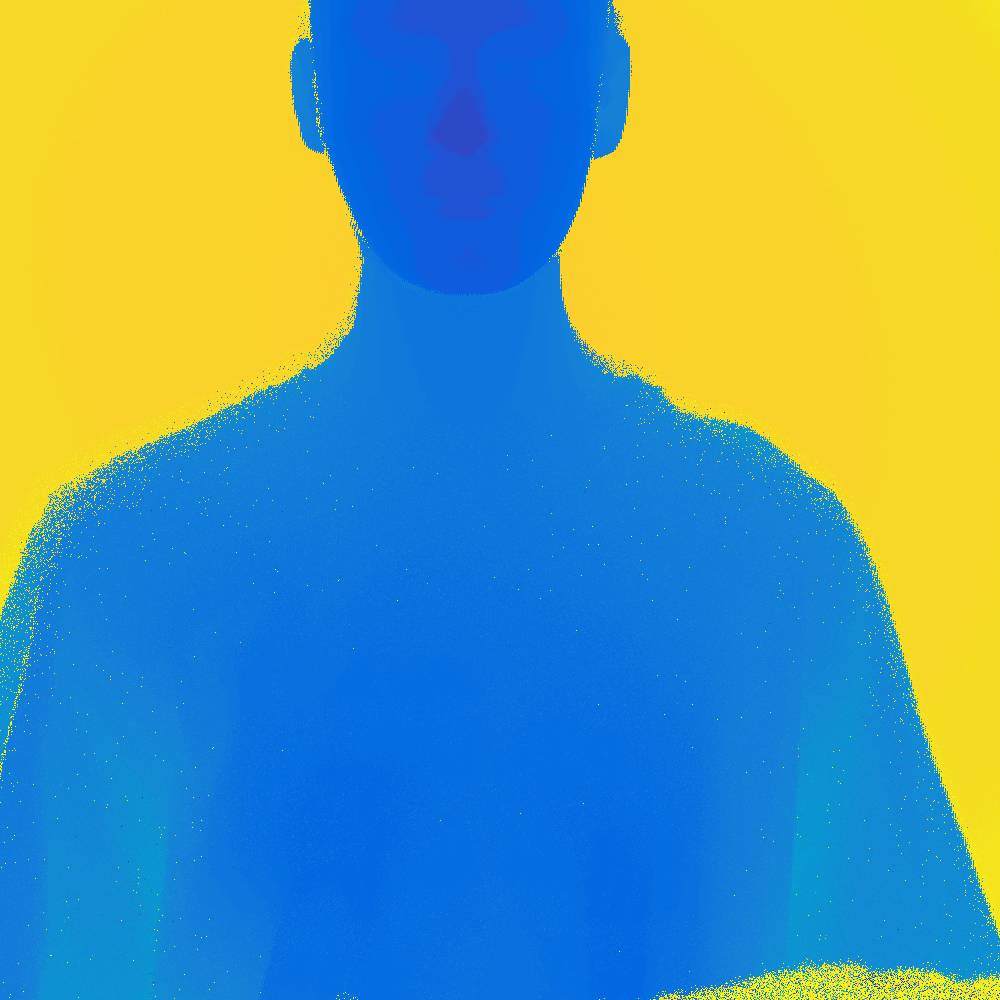}
        \small{\text{ }}
    \end{minipage}
    \begin{minipage}{ 0.14\linewidth}
        \centering
        \includegraphics[width=\textwidth]{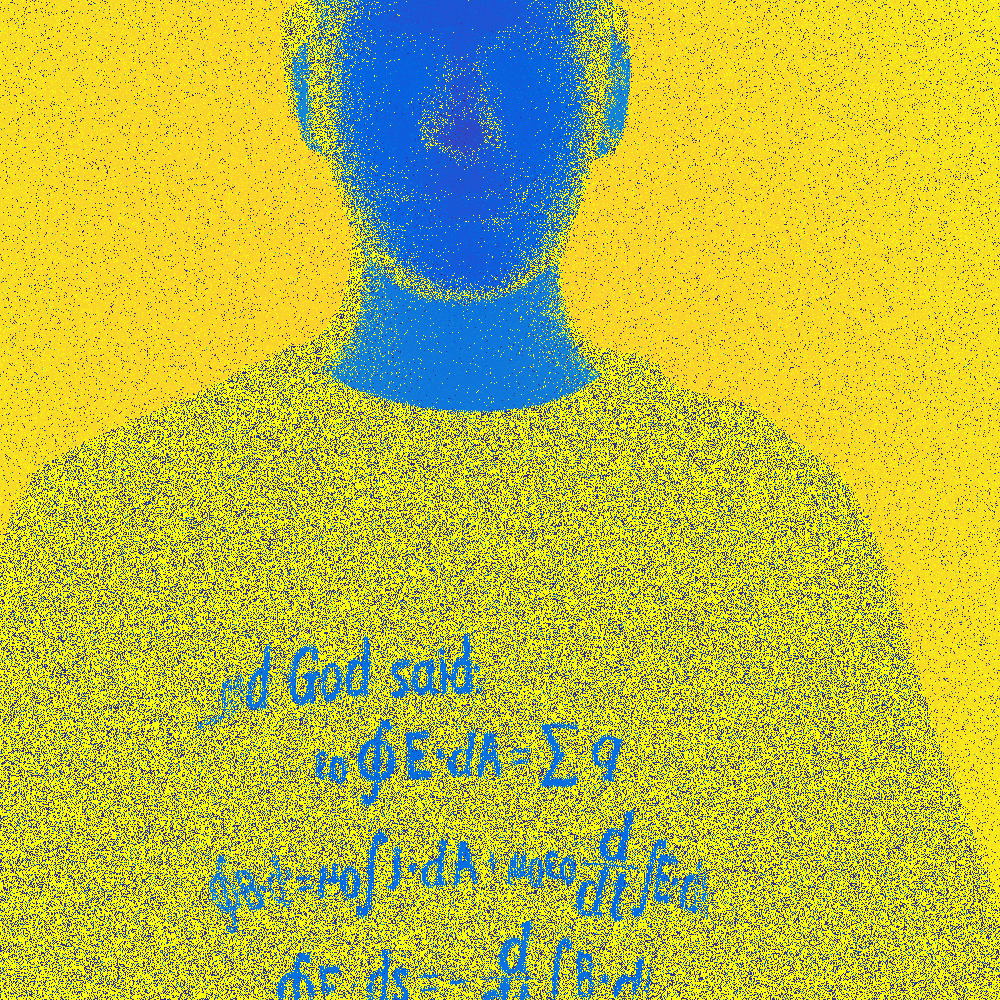}
        \small{\text{ }}
    \end{minipage} 
    \begin{minipage}{ 0.14\linewidth}
        \centering
        \includegraphics[width=\textwidth]{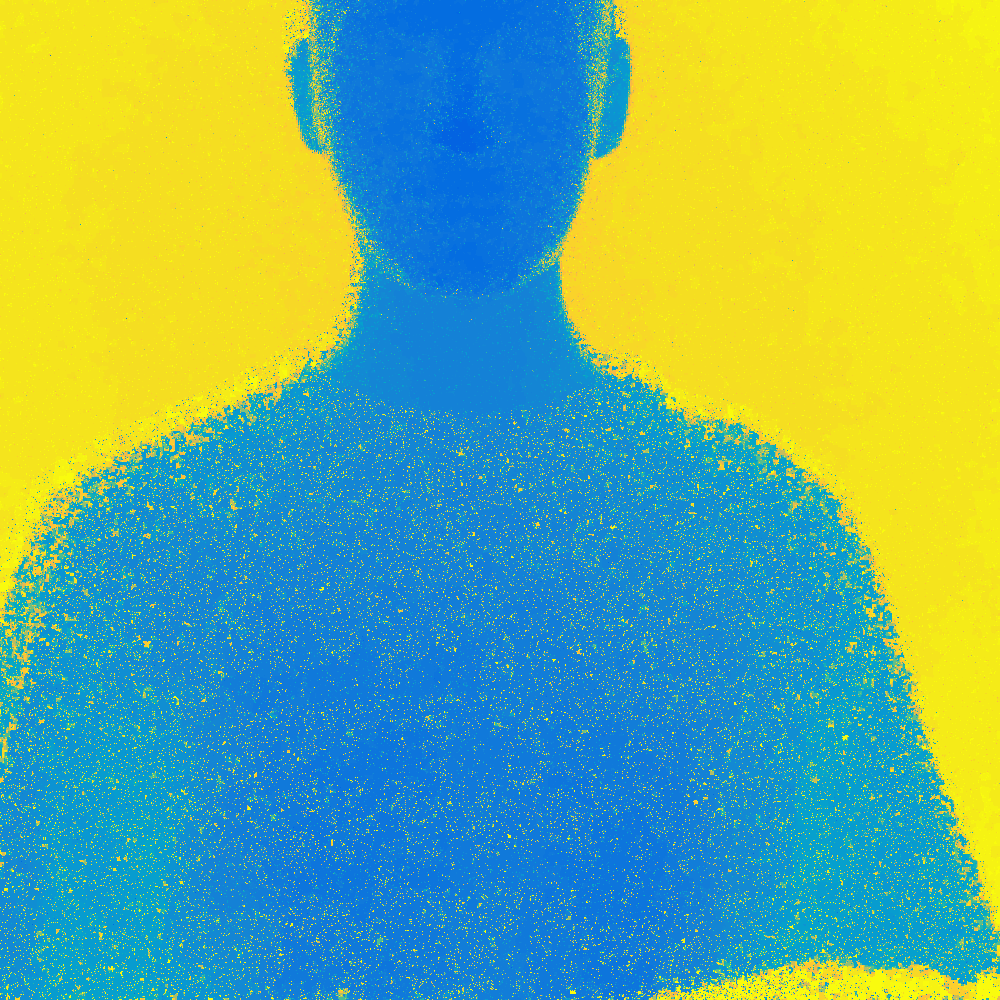}
        \small{\text{RMSE = 0.183 m}}
    \end{minipage} 
    \begin{minipage}{ 0.14\linewidth}
        \centering
        \includegraphics[width=\textwidth]{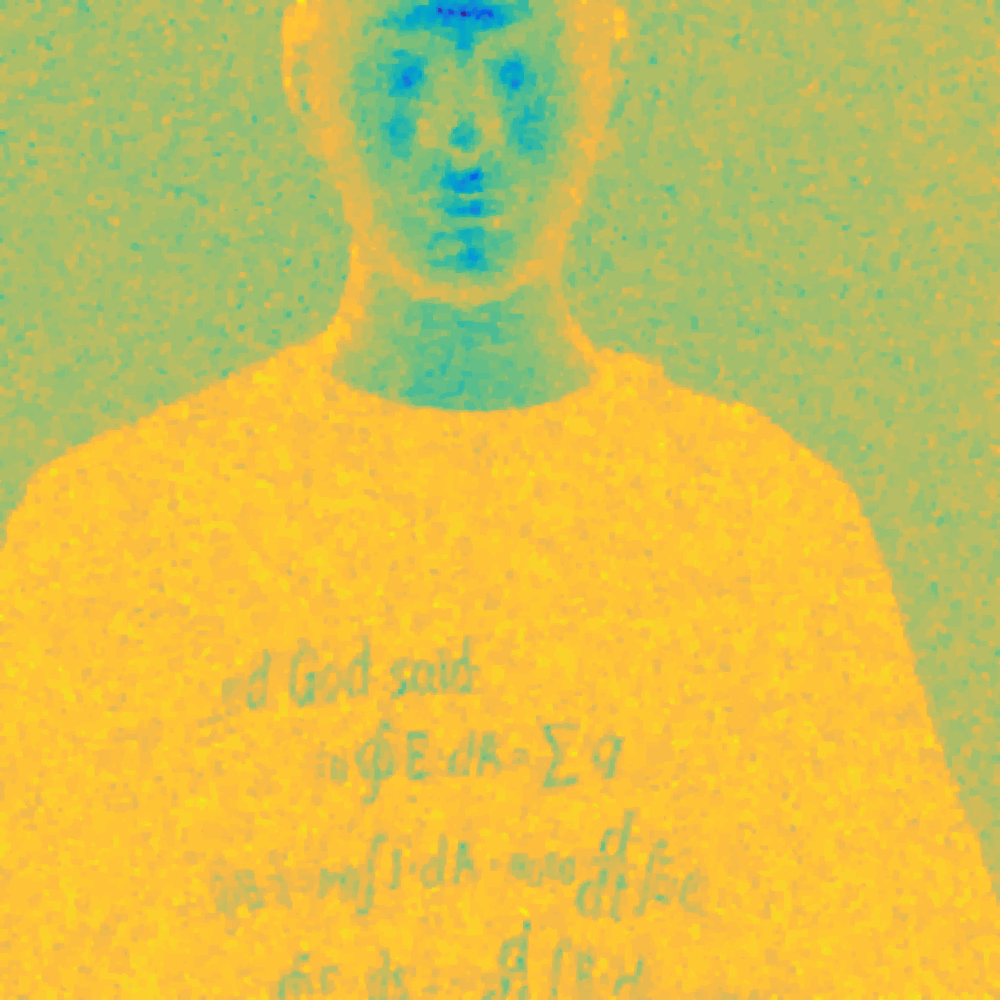}
        \small{\text{RMSE = 2.66 m}}
    \end{minipage} 
    \begin{minipage}{ 0.14\linewidth}
        \centering
        \includegraphics[width=\textwidth]{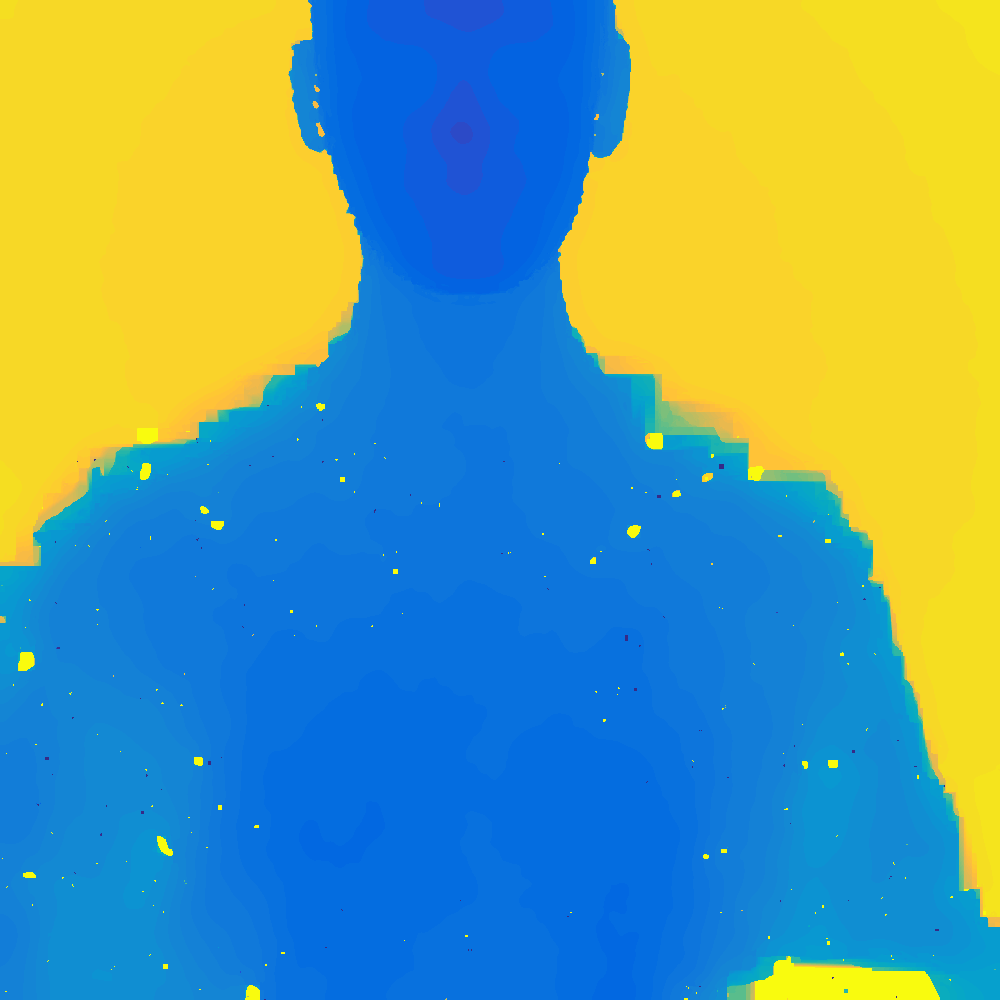}
        \small{\text{RMSE = 0.352 m}}
    \end{minipage} 
    \begin{minipage}{0.03\linewidth}
        \centering
        \includegraphics[ height=3.5\manheight]{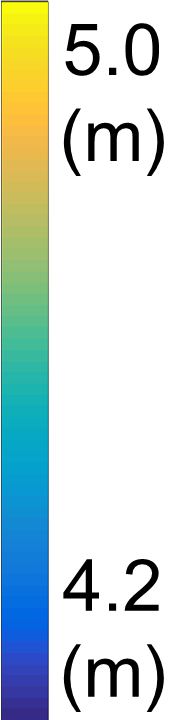}
        \small{\text{ }}
    \end{minipage}
    \begin{minipage}{ 0.14\linewidth}
        \centering
        \includegraphics[width=\textwidth]{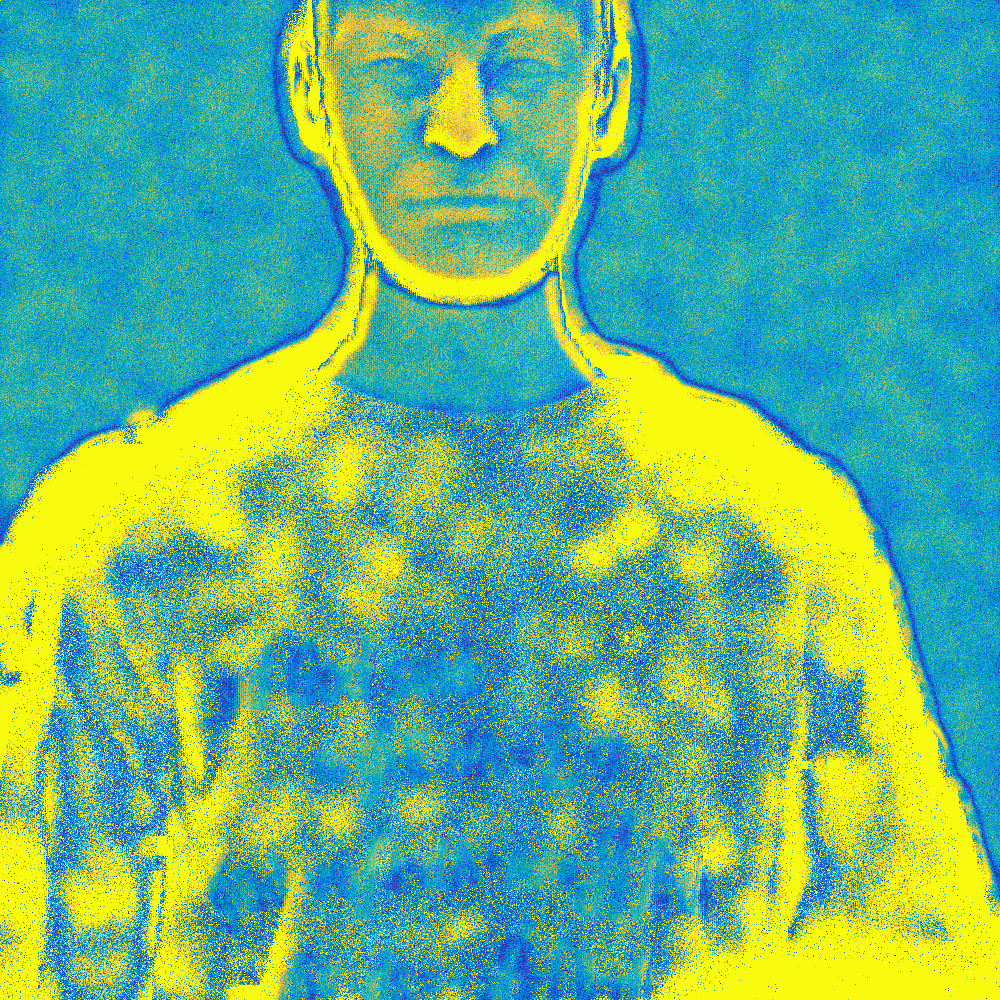}
        \small{\text{ }}
    \end{minipage} 
    \begin{minipage}{0.03\linewidth}
        \centering
        \includegraphics[ height=3.5\manheight]{Slide3}
        \small{\text{ }}
    \end{minipage}
\caption{Results of methods on the experimentally-acquired Mannequin data with about 4.05 signal detections per pixel and additional synthetically-generated noise to set the SBR at about 0.04. 
The error metrics are approximate, since the baseline LIDAR data is not exactly a ground truth for the scene.
Note that the depth estimates with the method of \cite{Shin2015} are completely out of range of the actual scene and are instead shown for the range of 5 to 8 meters.}
\label{fig:man_results}
\end{figure*}

We further evaluate the performance of our unmixing algorithm on the $1000 \times 1000$ pixel dataset of the Mannequin scene from \cite{github2015}, with results shown in Figure~\ref{fig:man_results}.
Baseline estimates were formed using conventional LIDAR processing on detection data from long acquisition times under constant conditions at SBR = 1.
The data was range-gated to capture the extent of the scene (4.2 to 6 meters), while limiting the influence of noise on the baseline estimates. 
Depth estimates were formed by applying the log-matched filter to the first 200 detections at each pixel.
Reflectivity estimates were formed by scaling the detection count by the number of illumination pulses required to reach 200 detections at each pixel.
Truncated photon-efficient datasets were created by using only the first 3000 illumination periods (300 $\mu$s per pixel), which resulted in 4.05 signal photons per pixel on average.
Additional background detections were synthetically generated as uniformly distributed detection times on $[0,\Tr)$ given a Poisson number of background detections.
Including the background detections already present in the data, the background rate was adjusted to set the SBR to 0.04 to match the simulated data.
For our algorithm, we use $\dspmax = 4$, $\tausp$ = 0.05, and $\tauFA = 0.01$.

The signal oracle processing was computed on the range-gated data, truncated to the first 3000 illumination periods.
Since the data was collected with ambient light injected into the scene, this data was not exactly noise-free.
This effectively sets SBR $\approx$ 8.3, so the oracle data represented much more favorable conditions than that for the other methods.
Still, there are certainly background detections present, as evidenced by the estimation of the mannequin  biased slightly to the mean acquisition range $\zmax/2$. 
This bias is most noticeable in the darkest areas such as the torso, where the local SBR is significantly lower than that for the entire scene.

Although there is little distinction when comparing the approximate MSE for the reflectivity estimates using all three methods, there is a clear advantage to using our method over that of {\Shin}. 
The results using the method of \cite{Shin2015} are far more smoothed with less contrast, making the text unreadable and the facial features harder to distinguish. 
Our method instead produces much clearer results, which compare very favorably to the oracle and baseline reflectivity estimates.

As in the simulations, the method of \cite{Shin2015} yields an estimate that is completely out of the range of the true scene. 
The ROM censoring is unable to handle such low SBR, so the entire estimate is dominated by noise, which yields an estimate very close to $\zmax/2$.
The resulting RMSE is then mostly an indication of how close the scene subject was to the middle of the imaging range: since the simulated scenes were positioned farther from the center of the scene, the RMSE measures were larger.
On the other hand, our unmixing method proves to be considerably more effective at handling the high levels of background.
In particular, the brightest regions (the wall, the mannequin's face, and the shirt around the text) have low absolute error. 
The largest errors occur as in the simulations at object boundaries and in the darkest regions, such as the several small patches of the mannequin's shirt that have considerable errors.
One reason that the mannequin unmixing fared worse than the simulations is the starker contrast between regions of high and low reflectivity, which makes the average number of signal detections less representative of the distinct regions. 
The mannequin's shirt is a very large region with very few signal detections, so a higher $\dspmax$ value and a higher average signal detection count were necessary to improve estimates.

\section{Conclusions}\label{sec:conclusion}
The conventional approaches to active imaging in significant ambient light are to increase either the acquisition time or the illumination power.
In many situations, neither solution is practically feasible. 
In the case of autonomous navigation, for instance, vehicle LIDAR systems need rapid depth acquisition using safe laser intensities and without draining the limited power resources.
The only possible approach is a photon-efficient solution, which can make accurate measurements from very little incident signal illumination, even when the ambient light levels are high.

Based on key observations of the probabilistic nature of the signal and background detection processes, a simple windowing approach yields an effective unmixing of the component detection processes. 
By setting cluster size requirements based on the easily-measured background rate, we ensure that the number of falsely accepted background detections is limited.
Remaining gaps where too few signal detections were collected can be effectively filled through the spatially-adaptive process of forming superpixels and aggregating detections within those regions.

Finally, a great benefit to our approach is the modularity of the algorithm, which leaves room for improvement with upgrades to the component blocks.
For the results presented in this paper, we perform only a few loops through the algorithm using well-tuned parameters that provide good results at low SBR.
While forming superpixels helps fill in values for many pixels with empty depth estimates, we still require some inpainting to fill in the rest.
An ideal approach would likely perform more iterations, incrementing $\Nsp$ by one until each pixel has a reasonable depth estimate.
A major factor preventing this Goldilocks approach for the just-right $\Nsp$ at each pixel is the computational cost of concatenating and windowing many large vectors of detections.
Better implementations of our code could take advantage of the embarrassingly parallel problem structure \cite{moler1986matrix} with more distributed or GPU-accelerated computations.
Additional approaches to possibly improve results include alternative superpixel definitions, such as the fast SLIC method \cite{achanta2012slic}, or regularizers such as Joint Basis Pursuit \cite{tosic2014learning} that take further advantage of correlations between depth and reflectivity images.

\appendix
In this appendix, we derive approximations for probabilities of clusters due to noise and due to signal.
The final expressions are simple enough to enable tabulation of $\Ncl(\Nsp \Nr B,\, \tauFA,\, \Twind)$.

\subsection*{Noise Clusters}
Since detection of noise photons is a homogeneous Poisson process,
given $n$ noise detections, the detection times $\{t_{i,j}^{(\ell)}\}_{\ell=1}^{n}$ are distributed as the order statistics of $n$ independent uniform random variables on $[0, \Tr)$ \cite{Snyder1991}.
Rescaling the set of ordered detections $0 < u_{(1)} < \dots < u_{(n)} < \Tr$ by $\Tr$ so they occur in the range $[0,1]$, the $k$th order statistic $U_{(k)}$ has the beta distribution $\beta(k,n+1-k)$.

According to \cite[Sect.~2.3]{david2004order}, the time difference $S_{\ell-k}^{(k)} = U_{(\ell)} - U_{(k)}$ between the $k$th and $\ell$th detections where $1\leq k < \ell \leq n$ is $\beta((\ell-k),n+1-(\ell-k))$, which is a beta-distributed random variable that depends only on the difference between the $\ell$ and $k$ and not on their particular values.

Recall that $\Ncl$ denotes the minimum number of detections needed in a window of size $\Twind$ to consider that window as having a cluster of detections.
To have $\Ncl$ detections in a window beginning at the $k$th detection, we must have $S_{(k+\Ncl-1)-k}^{(k)} = S_{\Ncl-1}^{(k)} < \Twind / \Tr$.

Now for pixel $(i,j)$ to not have any clusters, we need all candidate windows to not have clusters.
Since there are $n$ noise detections, any of the first $n-(\Ncl-1)$ detections may be followed
by $\Ncl - 1$ additional detections within an interval of $\Twind$
and thus these are candidates for the beginning of a cluster.
Then the probability of no clusters is 
\begin{align*}
    \Pr&[\text{no cluster at }(i,j)\,|\,N_{i,j}=n] \nonumber \\
    &=\Pr[\{\text{no cluster starting at detection 1}\}\cap \dots \nonumber \\
    &\quad \cap \{\text{no cluster starting at detection }(n-\Ncl+1)\}\\
    &\quad |\,N_{i,j}=n].
\end{align*}
The different candidate windows are overlapping and thus the intersected events above are not independent.
Making an independence assumption greatly simplifies the computation
and gives an approximation that is
supported by the numerical evaluations
shown in Figure~\ref{fig:noise_clust}:
\begin{align}
    \Pr&[\text{no cluster at }(i,j)\,|\,N_{i,j}=n] \nonumber \\
    &\approx (\Pr[\text{no cluster starting at detection 1}\,|\,N_{i,j}=n])^{n-\Ncl+1} \nonumber \\
    &= (1-\Pr[S_{\Ncl-1}^{(1)} < \Twind / \Tr\,|\,N_{i,j}=n])^{n-\Ncl+1}.
\end{align}
From this, we have that the conditional probability of a cluster satisfies
\begin{align}
    \Pr&[\text{cluster at }(i,j)\,|\,N_{i,j}=n] \nonumber \\
    &\approx 1-(1-\Pr[S_{\Ncl-1}^{(1)} < \Twind / \Tr\,|\,N_{i,j}=n])^{n-\Ncl+1}.
\end{align}
Finally, since $N_{i,j}$ is Poisson-distributed, we can approximate the unconditional probability of a cluster by
\begin{align}\label{eq:noise_cluster}
    \Pr&[\text{cluster at }(i,j)] \nonumber \\
    &\leq \sum_{n=\Ncl}^\infty
    \underbrace{\Pr[N_{i,j} = n]}_{\Poisson(\Nsp\Nr B)} \nonumber \\
    & \qquad \cdot \big (1-(1-\underbrace{\Pr[S_{\Ncl-1}^{(1)} < \Twind / \Tr\,|\,N_{i,j}=n]}_{\beta(\Ncl-1,n+1-(\Ncl-1))})^{n-\Ncl+1} \big).
\end{align}

\subsection*{Signal Clusters}
We would like to derive the probability of clusters due to signal in a similar way, but we operate under the assumption of a Gaussian pulse shape, and the order statistics for the normal distribution are not available in closed form.
Instead, we restrict ourselves to consider a cluster present only when $\Ncl$ signal detections occur in a window of length $\Twind$ centered at the true depth.
Since we omit other window positions, we obtain a lower bound for the probability of a cluster being present.

The detection time of a signal photon, shifted based on the true depth and divided by $\Tp$, is given by a standard normal random variable.  Denoting the standard normal CDF by $\Phi$,
we have the probability of any particular detection landing in the centered window as 
\begin{equation}
    \Pwind = \Phi \left (\frac{\Twind}{\Tp} \right)-\Phi \left (-\frac{\Twind}{\Tp} \right).
\end{equation}
Given $m$ signal detections,
the probability that exactly $k$ of them land in the centered window is
\begin{align}
    \Pr[\text{exactly }&k\text{ detections in centered window}\,|\,M_{i,j}=m] \nonumber \\
    &={m \choose k} (\Pwind)^k (1-\Pwind)^{m-k}.
\end{align}
The conditional probability of no signal cluster at $(i,j)$ is the probability of having fewer than $\Ncl$ of the $m$ detections in the window, which is
\begin{align}
    \Pr[&\text{no cluster in centered window}\,|\,M_{i,j}=m] \nonumber \\
    &=\sum_{k=0}^{\Ncl-1} {m \choose k} (\Pwind)^{k} (1-\Pwind)^{m-k}.
\end{align}
Finally, since $M_{i,j}$ is Poisson-distributed, the unconditional probability of a signal cluster at $(i,j)$ is bounded as
\begin{align}
    \Pr[&\text{cluster at $(i,j)$}] \nonumber \\
    &\geq \Pr[\text{cluster in centered window}] \nonumber \\
    &= \sum_{m=\Ncl}^\infty \underbrace{\Pr[M_{i,j} = m]}_{\Poisson(\Nr \eta \alpha_{i,j} S)} \nonumber \\
    &\quad \cdot \Bigg (1- \underbrace{\sum_{k=0}^{\Ncl-1} {m \choose k} (\Pwind)^{k} (1-\Pwind)^{m-k}}_{\binomial(m,\Pwind) \text{ cdf at }\Ncl-1} \Bigg).
\end{align}

\section*{Acknowledgment}
The authors thank Dongeek Shin for providing code beyond \cite{github_TCI2015} and several extraordinarily helpful conversations.
Computing resources provided by Boston University's Research Computing Services are gratefully appreciated.

\bibliographystyle{ieeetr}

\end{document}